\documentclass[10pt, table, twoside]{article}
\usepackage[a4paper, textwidth=17cm, textheight=24cm]{geometry}

\usepackage{setspace}

\usepackage{lineno,hyperref}
\usepackage{graphicx}
\usepackage[singlelinecheck=false]{caption}
\usepackage{subcaption}
\usepackage{multirow}
\usepackage{adjustbox}
\usepackage{tabularx}
\usepackage{float}

\usepackage{graphicx} 
\usepackage{chemformula} 
\usepackage{amssymb}
\usepackage{amsmath}
\usepackage{bm}
\usepackage{mathrsfs}
\usepackage{amsfonts}
\usepackage{upgreek}
\usepackage{textgreek}

\usepackage{booktabs} 
\usepackage{todonotes}
\usepackage{comment}
\usepackage{multirow}

\usepackage[style=numeric,giveninits,sorting=none,backend=biber, uniquename=false ,maxbibnames=30, maxcitenames=2, natbib=true]{biblatex}
\makeatletter
\newcommand*{\transpose}{%
  {\mathpalette\@transpose{}}%
}
\newcommand*{\@transpose}[2]{%
  \raisebox{\depth}{$\m@th#1\intercal$}%
}
\makeatother

\usepackage{authblk}

\begin{document}

\title{Lattice Discrete Particle Model (LDPM): Comparison of Various Time Integration Solvers and Implementations}

\author[1,14]{Erol Lale}
\author[2]{Jan Eliáš\thanks{Corresponding author: jan.elias@vut.cz}}
\author[1]{Ke Yu}
\author[3]{Matthew Troemner\thanks{Dr. Troemner performed the work relevant to Sec. 3.1 during his PhD studies at Northwestern University.}}
\author[2]{Monika Středulová}
\author[4]{Julien Khoury}
\author[5]{Tianju Xue}
\author[6]{Ioannis Koutromanos}
\author[7]{Alessandro Fascetti}
\author[1,14]{Bahar Ayhan}
\author[7]{Baixi Chen}
\author[8]{Giovanni Di Luzio}
\author[9]{Yuhui Lyu}
\author[10]{Madura Pathirage}
\author[4]{Gilles Pijaudier-Cabot}
\author[11]{Lei Shen}
\author[12]{Alessandro Tasora}
\author[13]{Lifu Yang}%
\author[5]{Jiawei Zhong}
\author[1]{Gianluca Cusatis}

\affil[1]{Northwestern University, Department of Civil and Environmental Engineering, Evanston, IL 60208, USA}
\affil[2]{Brno University of Technology, Faculty of Civil Engineering, Institute of Structural Mechanics, 
Brno, 60200, Czechia}
\affil[3]{Cusatis Computational Services Inc, Wilmette, IL, 60091, USA}
\affil[4]{Université de Pau et des Pays de l'Adour, CNRS, LFCR, Allée du Parc Montaury, Anglet, 64600  France}
\affil[5]{Department of Civil and Environmental Engineering, The Hong Kong University of Science and Technology, Clear Water Bay, Hong Kong, 999077, China}
\affil[6]{Department of Civil and Environmental Engineering, Virginia Tech, 
Blacksburg, VA 24061, USA}
\affil[7]{Department of Civil and Environmental Engineering, University of Pittsburgh, Pittsburgh, PA 15261, USA}
\affil[8]{Politecnico di Milano, Department of Civil and Environmental Engineering, Milan, 20133, Italy}
\affil[9]{Department of Civil Engineering, The University of Hong Kong, Hong Kong, 999077,China}
\affil[10]{Gerald May Dept. of Civil, Construction and Environmental Engineering, Univ. of New Mexico, Albuquerque, NM 87131, USA}
\affil[11]{Hohai University, College of Water Conservancy and Hydropower Engineering, Nanjing, 210098, China}
\affil[12]{University of Parma, Department of Engineering for Industrial Systems and Technologies, Parma, 43121, Italy}
\affil[13]{City University of Hong Kong, Department of Architecture and Civil Engineering, Hong Kong, 999077, China}
\affil[14]{Istanbul Technical University, Department of Civil Engineering, Istanbul, 34469, Turkiye}

\maketitle

\section*{Abstract}
This article presents a~comparison of various implementations of the Lattice Discrete Particle Model (LDPM) for the numerical simulation of concrete and other heterogeneous quasibrittle materials. The comparison involves the use of transient implicit and explicit solvers and steady-state (static) solvers  and implementations for Central Processing Unit (CPU) as well as Graphics Processing Unit (GPU). The various implementations are compared on the basis of a~set of benchmarks tests describing behaviors of increasing computational complexity. They include elastic vibrations, confined strain-hardening compressive response, tensile fracture, and unconfined strain-softening compressive response.  Metrics of interest extracted from the simulations include macroscopic stress versus strain responses, computational times, number of iterations, and energy balance error. Pairwise comparison of final crack patterns is provided through the correlation coefficient and normalized root mean square error of the crack opening vectors. Moreover, for the most numerically challenging case of unconfined compression with sliding boundary conditions, the stability of the strain-softening response is tested by perturbing the solutions as well as changing the convergence criteria and time step size. Attached to this paper is the complete input data of the benchmark tests; this will allow researchers to run the examples and compare them with their own implementations. In addition, most of the reported implementations are publicly available in open source packages.  

\section*{Keywords}
Lattice Discrete Particle Model; LDPM; softening; heterogeneity; fracture; inelasticity; explicit solver; implicit solver.

\section{Introduction}

The Lattice Discrete Particle Model (LDPM) simulates heterogeneous materials at the length scale of major heterogeneity, such as coarse aggregate pieces for concrete~\parencite{cusatis2011lattice1}, fine aggregate pieces for mortar~\parencite{ troemner2022multiscale},  or grains in rocks~\parencite{li2017multiscale, li2018multiphysics,ashari2017lattice}. By combining random particle placement according to a~given particle size distribution~\parencite{cusatis2011lattice1,YanTro-24} (Fig.~\ref{fig:LDPM2}a), Delaunay tetrahedralization of the particle center and an~associated custom tessellation, LDPM systems for generic geometrical shapes consists of a~set of three-dimensional polyhedral cells, each representing the simulated heterogeneity and a~layer of embedding matrix (Fig.~\ref{fig:LDPM2}b). Adjacent polyhedral cells interact through shared triangular facets which are supposed to represent the potential location of cracks and other nonlinear features of the response of materials under load (Fig.~\ref{fig:LDPM2}b). 

LDPM builds its formulation on the pioneering work on lattice~\parencite{HerHan-89,schlangen1992simple, SCHLA92, bolander1998fracture, BOL99} and 
particle~\parencite{BAZTAB90,CUND2,CUNDSTRA1,SERO1,ROOR1,ZUELE80} models for tensile fracture
that started in the late eighties. Such work was then extended to also simulate fracture in compression and the transition from strain softening to strain hardening for confined compression in the formulation of the Confinement Shear Lattice (CSL) model~\parencite{cusatis2006confinement,confinementlattice2}.

In the last two decades, LDPM has been adopted in a~variety of different applications, including, but not limited to, concrete quasi-static fracture and damage~\parencite{cusatis2011lattice2,LDPM-F1,ZhuPat-22}, concrete dynamic fracture~\parencite{smith2017numerical, troemner2022multiscale}, simulation of reinforced concrete elements~\parencite{lale2018homogenization,alnaggar2019lattice,BhaGom-21}, concrete deterioration and aging~\parencite{ASR-FULL,YanPat-21}. Furthermore, LDPM can be conveniently coupled with discrete models for the simulation of mass transport and heat transfer~\parencite{li2018multiphysics,shen2020multiphysics, pathirage2018multiscale,pathirage2018effect} for the solution of multiphysics poromechanics problems in which the mechanical behavior, particularly cracking, is fully coupled with the effects of pore pressure and temperature~\parencite{yin2024interprocess}. 

LDPM provides superior predictive capability over other continuum based models because it features, automatically, a~material length scale anchored to material heterogeneity. This allows simulation of nonlocal effects on strain-softening behavior without numerical artifacts or the computational complications of nonlocal gradient or integral models~\parencite{bavzant1994nonlocal, de1995comparison, DiLuzio2007, peerlings1996gradient, pijaudier1987nonlocal, HAVLASEK201672,lale2017isogeometric,PijKho24}.

Traditionally, LDPM has been implemented using \emph{explicit} time integration solvers. Specifically, the central difference scheme was successfully adopted in many different studies~\parencite{cusatis2011lattice2,lale2018homogenization}.  Explicit implementations avoid convergence issues typical for \emph{implicit} algorithms that always hamper the simulation efficiency in the presence of strain softening~\parencite{DeB87}. However, explicit schemes for the solution differential equations~\parencite{courant1967partial, zienkiewicz2000finite, bathe2006finite} are only conditionally stable. For a~typical LDPM system for concrete the stable time step is of the order of $10^{-7}$ sec. This limits the total simulation time to a~few seconds for most cases. This is of course not an~issue for the simulation of the response under highly impulsive loads whose duration is of order of magnitude $10^{-2}$ seconds. However, it definitely becomes an~issue for the simulation of quasi-static loading applications with duration in the order of several minutes and even more so for sustained loading lasting several decades. This has required various workarounds in which \emph{explicit} solvers are used to compute solutions relevant to a~much higher loading rate than the real one~\parencite{cook2007concepts, smith2017numerical, smith2014discrete}. Previous studies used ``real'' to ``simulation'' time maps  so that the constitutive equations would be calculated at the material response under the ``real'' strain rates~\parencite{yin2024interprocess}. Quasi-static simulations with artificially high loading rates  have been considered realistic if the ratio between kinetic energy, $E_{\mathrm{k}}$, and internal energy, $W_{\mathrm{int}}$, is below a~threshold on the order of $10^{-3}$. 

On the other hand, \emph{implicit} solvers are unconditionally stable, time integration of the response under quasi-static and sustained loading conditions can be obtained with much larger time steps. However, the convergence of the iterative algorithm at each step might become painfully slow or even fail, in case of a~response featuring multiple locations with a~strain softening response.  Another complication with implicit solver lies in the necessity to assemble some form of stiffness matrix and set the convergence criteria. \emph{Implicit} solver implementations for LDPM or similar models have been reported in the literature both in the steady state and transient regime~\parencite{doi:10.1061/JMCEA3.0000098,EliVor-15,EliVor20,jia2024efficient}. 

The main objective of this study is to provide a~comparison of the performance of multiple \emph{implicit} and \emph{explicit} solvers implemented in various different software packages to solve the LDPM equations of motion. The comparison is pursued on five typical tests of increasing computational complexity: (i) free elastic vibrations; (ii) confined compression test, which features a~macroscopic hardening response and a~diffused material inelastic behavior; (iii) three-point bending test of notched samples and (iv) direct tension of dogbone type specimen featuring macroscopic softening response and the propagation of one macroscopic fracture; and (v) unconfined compression test, characterized by a~macroscopic softening response and the propagation of a~complex pattern of cracks.

 This paper demonstrates that \emph{implicit} solvers can be successfully used for LDPM simulations even in the presence of extremely complex nonlinear behavior with large number of cracks developing during the loading process. The paper shows in which situations the \emph{implicit} time integration becomes advantageous over \emph{explicit} time integration and when the opposite is true. In addition, this paper and its associated data serve as a~tutorial for the verification of future LDPM implementations. 

\section{LDPM Governing Equations}

\begin{figure}[t!]
     \centering
     \includegraphics[width=1.0\textwidth]{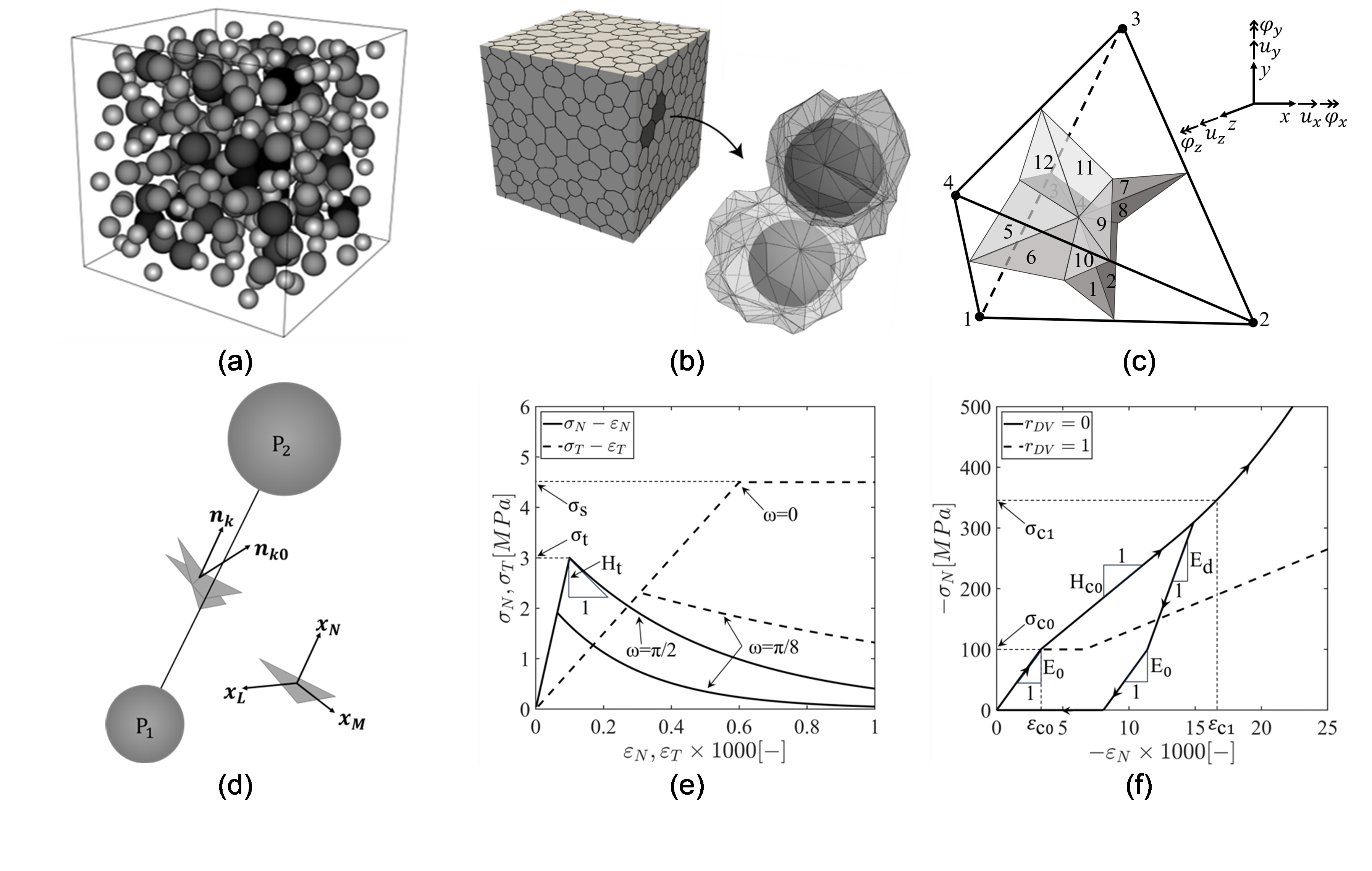}
     \caption{(a) LDPM particles in a~cubic geometry following placement procedure; (b) set of two LDPM polyhedral cells composed of a~single particle and their surrounding facets; (c) set of four LDPM particles and associated facets; (d) original and projected LDPM facets; (e) typical traction versus strain curves at the LDPM facet level; (d) typical normal traction versus normal strain curves in compression.}
     \label{fig:LDPM2}
\end{figure}

LDPM adopts rigid body kinematics to define the measure of strains at the interface of adjacent polyhedral cells. These strain measures (Eq. \ref{eq:strain1}) are computed at the centroid of each facet $k$  through a~displacement jump, $\left[\kern-0.15em\left[ {{{\mathbf{u}}}}  \right]\kern-0.15em\right]_{k}$, calculated from translations $\mathbf{u}$ and rotations $\bm{\uptheta}$ (degrees of freedom, DoFs) of adjacent nodes $I$ and $J$, operating in a~projected facet system of reference to avoid non-symmetric behavior in pure shear~\parencite{cusatis2011lattice1,HIGH_MP}. One has
\begin{align}
{\mathbf{e}}_k = \frac{1}{l_k}{{\mathbf{P}}^{\transpose}_k} \cdot{\left[\kern-0.15em\left[ {\mathbf{u}} 
 \right]\kern-0.15em\right]_{k}}
 \label{eq:strain1}
\end{align}
where ${\left[\kern-0.15em\left[ {\mathbf{u}} 
\right]\kern-0.15em\right]_{k}} = {{\mathbf{u}}_J} + {{\boldsymbol{\uptheta}}_J} \times {{\mathbf{c}}_k^J} - {{\mathbf{u}}_I} - {{\boldsymbol{\uptheta}}_I} \times {{\mathbf{c}}_k^I}$, $l_k$ is the length of the tetrahedron edge connecting the particles $I$ and $J$, ${\mathbf{P}}_k=[\mathbf{n}_k~\mathbf{m}_k~\mathbf{l}_k]$, and $\mathbf{n}_k$, $\mathbf{m}_k$ and $\mathbf{l}_k$ define the appropriate orthonormal vectors of the local reference system (Fig. \ref{fig:LDPM2}d). The normal vector $\mathbf{n}_k$ aligns with the straight connection of nodes $I$ and $J$, the tangential vectors can be chosen arbitrarily.  The vector $\mathbf{c}^I_k$ points from the node $I$ to the centroid of the $k$th facet. The LDPM facet strains correspond exactly to the local projection of a~strain tensor on the local reference system of the facet~\parencite{HIGH_MP}.

Equilibrium is enforced by the linear and angular momentum balance equations of each polyhedral cell as in Equations~\eqref{eq:balance}
\begin{subequations}
\label{eq:balance}
\begin{align}
\sum\limits_{k \in {\mathcal{F}_I}} A_k \mathbf{P}_k\cdot\mathbf{t}_k + V_I \mathbf{b} &= \mathbf{M}_u^I \cdot \ddot{\mathbf{u}}_I  + \mathbf{M}_{\theta}^I \cdot \ddot{\bm{\uptheta}}_I
\label{eq:balance_linear}
\\
\sum\limits_{k \in {\mathcal{F}_I}} A_k \mathbf{c}_k^I \times \left(\mathbf{P}_k\cdot\mathbf{t}_k\right) + V_I \mathbf{r}_I\times \mathbf{b} &= \mathbf{I}_{u}^I \cdot\ddot {\mathbf{u}}_I  + \mathbf{I}_{\theta}^I \cdot\ddot{\mathbf{\bm{\uptheta}}}_I
\label{eq:balance_angular}
\end{align}
\end{subequations}
where $\mathcal{F}_I$ contains all facets of a~polyhedral cell $I$, $A_k=A_{0k}\mathbf{n}_k\cdot\mathbf{n}_{k0}$ is the projected area of a~facet orthogonal to the corresponding tetrahedron edge, $\mathbf{n}_{k0}$ is the true normal to the facet plane, $V_I$ is the cell volume, $\mathbf{r}_I$ is the vector from the particle center to the cell centroid,  $\mathbf{b}$ is the external body force (constant over the overall solid volume of interest), $\mathbf{t}_k$ is the traction vector in local reference system, and $\mathbf{M}^I_u$, $\mathbf{M}^I_{\theta}$, ${\mathbf{I}}^I_u$, and ${\mathbf{I}}^I_{\theta}$ are inertia tensors of the cell. 


The LDPM governing equations are then completed by a~set of vectorial constitutive equations relating tractions and strains: ${\mathbf{t}}_k=\mathbf{T}({\mathbf{e}}_k, {\mathbf{s}}_k, \mathbf{p})$ where ${\mathbf{s}}_k$ is a~vector of internal variables and $\mathbf{p}$ is a~vector of model parameters. In the elastic regime, one has ${\mathbf{t}}_k={\mathbf{E}}\cdot{\mathbf{e}}_k$ where ${\mathbf{E}}=E_0 \textrm{\bf{diag}} (1\;\alpha\;\alpha)$ where $E_0$ and $\alpha$ are model parameters known as the effective normal modulus and shear-normal coupling parameter. LDPM nonlinear, inelastic constitutive equations have been successfully formulated and validated for concrete and other quasi-brittle materials in previous work, to which interested readers are directed for additional information. \ref{appendix:material model} reports the nonlinear constitutive equations formulated in Ref.~\parencite{chiara} and
adopted in this work. Figures \ref{fig:LDPM2}e and f show examples of the facet traction versus strain response according to the adopted LDPM constitutive equations for tension-shear and compression, respectively. The variable $\omega$ appearing in Figure \ref{fig:LDPM2}e is defined by the expression $\tan(\omega)=e_N/[\alpha\left(e_M^2+e_L^2\right)]^{1/2}$; $\omega=0$ simulates pure-shear; $\omega=\pi/2$ simulates pure tension; and $0< \omega<\pi/2$ corresponds to a~mixed-mode loading condition.

\section{Numerical Implementations}

LDPM can be conveniently implemented in traditional finite element codes with reference to the underlying Delaunay tetrahedral mesh. Each LDPM tetrahedral element is composed of 4 nodes with rotational and translational degrees of freedom and is tessellated with 12 facets (Fig.~\ref{fig:LDPM2}c). Various existing publications~\parencite{cusatis2011lattice1, cusatis2011lattice2, LDPM_report} present a~detailed derivation of the stiffness matrix of the LDPM tetrahedral element, the mass matrix, and the force vector, which are omitted here for brevity. All numerical implementations presented in this paper are based on a~Lagrangian formulation under the hypothesis of small strains, small displacements, and small rotations.

After assemblage of all LDPM tetrahedral elements, the general form of the equations of motion for LDPM system can be written as
\begin{align}
   \mathbf{M} \cdot \ddot{\mathbf{q}} + \mathbf{f}_{\text{int}}(\mathbf{q}) = \mathbf{f}_{\text{ext}} 
\end{align}
where $\mathbf{M}$, $\mathbf{f}_{\text{int}}$, $\mathbf{f}_{\text{ext}}$, and $\mathbf{q}$ denote the mass matrix, the internal forces, the external forces, and the vector of degrees of freedom (combined $\mathbf{u}$ and $\bm{\uptheta}$),  respectively. In the elastic case $\mathbf{f}_{\text{int}}=\mathbf{K} \cdot \mathbf{q}$, where $\mathbf{K}$ is the stiffness matrix (Hessian of potential energy). The mass matrix can be consistent~\parencite{archer1963consistent, cusatis2011lattice1} or diagonalized, which is obtained simply by retaining the diagonal terms of the consistent mass matrix. The off-diagonal terms are omitted in the lumped matrix because their addition to the diagonal terms is inappropriate since they have different units.

\subsection{LDPM Mesh Generation}
All simulations performed in this study relied on the LDPM geometric input provided by NU-FreeCAD\footnote{primary contacts: Erol Lale, lale@itu.edu.tr, and Matthew Troemner, mtroemner@gmail.com}. FreeCAD is an~open-source parametric 3D CAD modeler suitable for a~variety of engineering and design tasks. Its modular construction allows operation on Windows, macOS, and Linux, and compatibility with numerous file formats. Its Python scripting integration provides powerful customization and automation possibilities, making FreeCAD a~viable backend tool for creating specialized workflows essential in research and development environments~\parencite{FreeCADDocumentation}. This Python integration also broadens its utility, catering to users ranging from beginners to experts~\parencite{FreeCADTutorial}. The Python back-end has been exploited for the purposes of model development of LDPM simulations. Through various Python scripts, the particle placement, meshing, and tessellation procedures outlined in Ref.~\parencite{cusatis2011lattice1} have been recreated and generalized. The NU-FreeCAD Preprocessor is distributed freely\footnote{\url{https://github.com/Computational-Mechanics-Material-Models/chrono-preprocessor}} under the BSD 3-Clause License.

Upon opening an~NU-FreeCAD enabled version of FreeCAD, the user is prompted with an~additional Workbench for LDPM functionality. All necessary inputs, as described in~\parencite{cusatis2011lattice1} for traditional concrete, include water-to cement ratio, $w/c$, cement content, $c$, maximum aggregate size, $d_a$, minimum particle size, $d_0$ and aggregate sieve curve. The aggregate sieve curve can be provided as a~piecewise linear curve~\parencite{YanPat-22} or by the fitted exponent of the Fuller curve.

The built-in utilities perform simple meshing procedures, while the more sophisticated LDPM-specific algorithms for mesh-construction were developed ad-hoc. Upon completion, all meshes can be visually displayed in the FreeCAD graphic interface while also output to files. For the solvers with built-in compatibility, users can also use the FreeCAD FEA utilities to apply boundary constraints, loads, and other simulation conditions.

LDPM polyhedral cell structures, tetrahedral mesh, and corresponding facet data (including facet nodes, normal/tangential components, projections, areas, etc.) are provided as output files that can be read and interpreted by the different solvers. 

\subsection{Explicit and Implicit Solvers}
 The following implementations of LDPM are selected for comparison in the present paper: ABAQUS Explicit via a~VUEL subroutine~\parencite{troemner2022multiscale}, Project Chrono~\parencite{tasora2016chrono}, JAX~\parencite{xue2023jax}, Open Academic Solver (OAS)~\parencite{EliCus22}, CAST3M~\parencite{verpaux2millard, PatThi-22}, FEMultiPhys~\parencite{koutromanos2024femultiphys} and Julia LDPM \parencite{chen2025stochastic, zhu2026lattice, chen2026displacement, jia2024efficient}.
The implementation of LDPM in these software packages, which features implicit and explicit solvers, is described in details in Ref.~\parencite{LDPM_report}. 

The implicit solvers all adopt unconditionally stable time integration schemes and they check convergence with various convergence criteria as highlighted in the following sections and discussed in detail in Ref.~\parencite{LDPM_report}. This makes it quite difficult to set time steps and convergence tolerances across the different solvers to make the comparison meaningful. 

Hence, it was decided to allow the different software developers to set those parameters based on their experience and knowledge of the specific implementation with the only requirement that all the simulations had to converge with less than 50 iteration most of the time.  However, a~higher number of iterations and possibly also acceptance of non-converging results was allowed if occurring at a~limited number of steps. 

It is worth noting that explicit solvers do not require setting any simulation-level parameters since they are not based on any convergence criterion and the time step is dictated by their stability condition.

Table~\ref{tab:Adopted-Software} reports basic information relevant to the adopted solvers. Table~\ref{tab:Implicit-Solvers} then summarizes implicit solvers and their settings related to Newton-Raphson iterative search for equilibrium.

\subsubsection[ABAQUS Explicit]{AE: ABAQUS Explicit\footnote{primary contact: Erol Lale, lale@itu.edu.tr}}

For streamlined simulations and integration with existing finite element and interaction libraries, the Lattice Discrete Particle Model was implemented through the ``VUEL'' user element subroutine in the commercial solver ABAQUS Unified FEA with explicit time solver~\parencite{troemner2022multiscale}  based on the central difference algorithm \cite{abaqus2016_theory}. Once imported, the LDPM mesh may be manipulated in the same manner as an~ABAQUS-native mesh, including the imposition of boundary conditions, constraints, interactions, and other features (e.g. mass scaling). It is worth mentioning that by using the built-in ABAQUS subroutines to govern the underlying LDPM calculations, the simulations may be fully parallelized. Thus, any LDPM domain can be decomposed in the same manner as any FEA decomposition.

\subsubsection[Project Chrono]{CI: Project Chrono\footnote{primary contact: Erol Lale, lale@itu.edu.tr}}
Chrono is an~open-source multi-physics engine designed to simulate dynamics of complex and large-scale systems. It is written in \texttt{C++} and designed to be modular and extensible. Chrono is particularly well-suited for simulating systems that involve frictional contact, multibody dynamics, fluid-solid interaction, granular dynamics, and finite element analysis. Chrono 
can be used for a~wide range of applications in engineering and science~\parencite{tasora2016chrono} and 
is publicly available \footnote{\url{https://github.com/projectchrono/chrono}} under the BSD-style license. The Chrono LDPM implementation can be found in the chrono-concrete branch \footnote{\url{https://github.com/Computational-Mechanics-Material-Models/chrono-mechanics}}.

Chrono implements many implicit and explicit time integration solvers, including 2$^{\mathrm{nd}}$ order Implicit time integrators such as  Hilber-Hughes-Taylor (HHT)~\parencite{HilHug-77}, Newmark methods~\parencite{doi:10.1061/JMCEA3.0000098} as well as 1st order implicit techniques such as linearized Euler and projected Euler methods~\parencite{tasora2016chrono}.

{\bf Hessian factorization and convergence criteria}: Numerical examples presented here are simulated using Chrono-V8, where the system matrix is updated and factorized once per time step (at the first iteration); however, the initial elastic stiffness is always used. Recently, Chrono released a~newer version V9, which can optionally update and factorize the system matrix only once at the beginning of the analysis. Chrono V8 employs two different sets of convergence criteria depending on the integration mode: ``acceleration'' or ``position''. Convergence is declared if either the residuals are sufficiently small or the state update is sufficiently small. For this paper, we selected the position mode, which checks convergence using the weighted root mean square (WRMS) method, as $\lVert \Delta u \rVert_{WRMS} < 1$. For WRMS tests, the threshold is always set to 1, since the error is internally normalized against the user-defined relative and absolute tolerances via weights calculated as $w_i = 1 / (r_{\text{tol}} × |u_i| + a_{\text{tol}})$. Here, $u_i$ denotes the degree of freedom (DoF, displacement and rotation), and $\Delta u_i$ represents its increment within the iteration. All simulations presented in this study adopted the HTT method with $\alpha=-0.05$, $\gamma=1/2-\alpha$ and $\beta=1/4 \left( 1 -\alpha \right)^2$ with relative and absolute tolerances equal to $r_{\text{tol}}=10^{-4}$ and $a_{\text{tol}}=10^{-6}$, respectively.

\subsubsection[Cast3M]{CA: Cast3M\footnote{primary contact: Gilles Pijaudier-Cabot, gilles.pijaudier-cabot@univ-pau.fr}}

Cast3M is a~computer code developed for (i) analysis of structures using the finite element method and (ii) analysis of flow through computational fluid dynamics. The code was originally developed by the Modeling of the Systems and the Structures Department of the French Atomic Energy Commission (CEA). Currently, Cast3M is a~general program for the analysis of linear elastic problems in statics and dynamics (vibration, eigenvalue analysis), thermal and heat transfer problems, nonlinear problems (in both geometry and material), step by step dynamic problems such as contact problems, etc. Unlike many other codes, Cast3M does not work as a~black box, but gives the user a~set of operators that can be combined to formulate and solve complex problems. This software uses a~high-level command language, called Gibiane, which allows the user to manipulate data and results as objects. The programming language for operators is ESOPE which can be seen as FORTRAN 77 of higher level with dynamic memory allocation. Cast3M is a~complete system, integrating not only the solver, but also functions for pre-processing and post-processing. 

The Lattice Discrete Particle Model was implemented in ESOPE. Although ESOPE allows for a~transient solution with consistent and lumped mass matrices by using the average acceleration method
, for this study, only the steady-state (static) solver was employed.

{\bf Hessian factorization and convergence criteria}: Cast3M uses an~initial stiffness which is computed and factorized only once at the beginning of simulation. The convergence criterion requires the $\ell_2$ norm of residual forces normalized by the $\ell_2$ norm of external forces to be below the threshold $10^{-4}$.

\subsubsection[Open Academic Solver (OAS)]{OA: Open Academic Solver\footnote{primary contact: Jan Eliáš, jan.elias@vut.cz}}

Open Academic Solver (OAS) is an~open source software publicly available\footnote{\url{https://gitlab.com/kelidas/OAS}} under the GPLv3 license. The core is written in \texttt{C++} and utilizes the object-oriented code structure. It originated as a~solver for steady-state mechanical problems, primarily focused on discrete mesoscale material representation. The early work combined spatial material randomness described by random field fluctuations in material properties with fracture~\parencite{EliVor-15,EliVor20}. A~substantial effort was devoted to allow an~adaptive refinement of the discretization~\parencite{Eli16,MasKve-23}. The code was later modified to incorporate transient solvers, both explicit and implicit, and multiphysical tasks~\parencite{EliCus22,MasKve-23}. A~crucial OAS feature is the ability to recursively embed sub-models allowing its use for FE$^2$ (or nested) homogenization~\parencite{EliYin-22,EliCus25,EliCus26}. The solver contains standard finite elements and lattice particle elements, such as LDPM tetrahedral elements and CSL two-node edge elements.

The OAS simulations presented here used the implicit generalized-$\alpha$ method of~\textcite{ChuHul93}, which is a~general method with four parameters $\alpha_m$, $\alpha_f$, $\gamma$ and $\beta$. For $\alpha_m=0$ one obtains the HHT method~\parencite{HilHug-77}, for $\alpha_f=0$ it becomes the WBZ method~\parencite{WooBoss-80}, setting $\alpha_m=\alpha_f=0$ results in the Newmark method~\parencite{doi:10.1061/JMCEA3.0000098}. Optimal behavior (second-order accuracy, damped high frequencies and undamped low frequencies) is obtained by setting $\alpha_m = \frac{2\rho_{\infty}-1}{\rho_\infty+1}$, $\alpha_f=\frac{\rho_\infty}{\rho_\infty+1}$, $\gamma = \frac{1}{2} - \alpha_m + \alpha_f$, and $\beta = \left(\frac{ 1 - \alpha_m + \alpha_f }{2}\right)^2$. The spectral radius $\rho_\infty$ was set at 0.8.

{\bf Hessian factorization and convergence criteria}: OAS uses in all the simulations in this paper an~initial stiffness which is computed and factorized only once at the beginning of simulation. The convergence criteria are adopted according to Ref.~\cite[page 332]{BelLiuMor00}, all with the convergence threshold $10^{-4}$. They are the $\ell_2$ norm of (i) residuals and (i) DoF increments normalized by the maximum of external, internal and inertia forces norms or the norm of total DoF values, respectively. The third convergence check computes the energy of residual forces on DoF increments and normalizes is by the total external, internal, or kinetic energy.

\subsubsection[\textit{FEMultiPhys} code]{MP: \textit{FEMultiPhys} code\footnote{primary contact: Ioannis Koutromanos, ikoutrom@vt.edu} \label{sec:MP}}
\textit{FEMultiPhys} is a~general-purpose, research-oriented finite element analysis program for the simulation of time-dependent coupled problems, including mechanical deformation, heat transfer, mass transport in porous media, and chemical advection/diffusion/reaction for multiple species. \textit{FEMultiPhys} is primarily tailored for analyses at the structural component and system level. The program is available for free use and can be downloaded through the \textit{DesignSafe} cyberinfrastructure of the National Science Foundation~\parencite{koutromanos2024femultiphys}.

The LDPM implementation in \textit{FEMultiPhys} is largely based on the original LDPM formulation as presented in Ref.~\parencite{cusatis2011lattice1}. The only major difference in the \textit{FEMultiPhys} version is in the procedure to obtain a~diagonal mass matrix. Specifically, the mass matrix is established assuming lumped (point) masses and an~isotropic (i.e., diagonal) rotatory inertia tensor for each particle. The latter assumption means that the diagonal rotatory masses corresponding to any Cartesian triad of axes are identical. The translational mass of each particle is first calculated, treating the LDPM tetrahedra as continuum regions. The total mass of each tetrahedron is taken as the product of the undeformed volume times the (constant) average mass density of the concrete, and then equally distributed to the four vertex particles. Then, the value of each diagonal element $I_m$ in the rotatory inertia tensor is calculated by treating the particle as a~solid sphere with diameter equal to the diameter of the particle $d_p$:
$I_m = m d_p^2 /10$. This approach may entail some advantages, for example, for large-rotation computations, as it leads to a~diagonal isotropic rotatory mass matrix. 

The LDPM computations in each step are vectorized; that is, each individual instruction (e.g., computation of facet normal strain) is performed simultaneously for a~block of facets. 

The simulations presented in this paper are based on the implicit dynamic solver of the code, using an~HHT time-marching scheme~\parencite{HilHug-77}. The solution proceeds in an~incremental-iterative fashion, and the iterations employ a~Newton-Raphson algorithm based on the initial tangent stiffness matrix. Convergence is established whenever the norm of the residual (unbalanced force) vector becomes less than a~user-specified tolerance.
Both absolute-value and relative-value criteria are available; the latter is adopted herein. The algebraic computations for the inversion (more accurately, factorization) of the stiffness matrix and the associated back-substitution operations for implicit solution employ the parallel direct sparse solver \textit{PARDISO}, as implemented in the Intel \textit{oneAPI Math Kernel Library} (\parencite{IntelMKL}).

{\bf Hessian factorization and convergence criteria}: Also FEMultiPhys uses the initial elastic stiffness and factorize it only once at the beginning of the simulation. For the convergence assessment,  the $\ell_2$ norm of residual forces normalized by the norm of internal forces is compared to the threshold $10^{-3}$.

\subsubsection[JAX-LDPM]{JA: JAX-LDPM\footnote{primary contact: Tianju Xue, cetxue@ust.hk}}
JAX-LDPM is an~open-source software\footnote{\url{https://github.com/tianjuxue/jax-ldpm}} in active development by researchers from the Hong Kong University of Science and Technology.  Unlike typical scientific computing libraries developed in compiled languages such as \texttt{C/C++} or \texttt{Fortran}, JAX-LDPM is a~pure \texttt{Python} library accelerated with just-in-time compilation techniques.

JAX-LDPM is based on JAX~\parencite{jax2018github}, a~growing Python library for general purpose high-performance numerical computing with a~focus on deep learning, developed and maintained by Google. As a~new AI framework, JAX is intended to replace TensorFlow as the new underpinning for some of the Google's products, due to its excellent flexibility and fast performance~\parencite{takamoto2022pdebench}. In addition to its recognizable success in machine learning research, JAX has proven to be a~powerful building block for high-performance scientific computing tasks, including computational fluid dynamics~\parencite{bezgin2023jax,kochkov2021machine}, structural dynamics~\parencite{xue2023learning}, finite element analysis~\parencite{xue2023jax}, structural shape optimization~\parencite{wu2023framework}, etc.
The same code naturally runs on the CPU and GPU, depending on the availability of hardware. 
Since JAX itself is a~deep learning framework, JAX-LDPM has access to all state-of-the-art deep learning models provided in the JAX ecosystem. JAX-LDPM uses the explicit central difference method for time integration \cite{abaqus2016_theory, bathe2006finite}.
The simulations presented in this study were performed with a~lumped mass matrix on a~GPU architecture. 

\subsubsection[Julia LDPM (JuLDPM) package]{JU: Julia LDPM (JuLDPM) package\footnote{primary contact: Alessandro Fascetti, fascetti@pitt.edu}}

The Julia LDPM (JuLDPM) package was developed and is currently maintained at the University of Pittsburgh as a~research-oriented tool for simulating coupled mechanical and transport phenomena in concrete~\parencite{fascetti2022stochastic}, geomaterials~\parencite{wang2024three}, and fiber-reinforced polymers~\parencite{fascetti2016web}. 

The LDPM implementation follows the original work of~\textcite{cusatis2011lattice1} for the definition of the computational framework. The package is written entirely in the Julia programming language, which is a~dynamic language developed for scientific computing applications~\parencite{bezanson2012julia}. The original implementation of the method is described in detail in the previous work of the authors~\parencite{fascetti2018lattice}. Recent advancements introduced in~\parencite{jia2024efficient} enable the use of a~static solver for the LDPM governing equations. 
Although the LDPM kinematics is maintained identical to the original formulation, several local modifications of the constitutive laws were introduced to render the material response continuous for any combination of traction and strain measures at the lattice level. However, for the sake of a~direct comparison with other implementations used in this study, all JAX-LDPM simulations were performed using the original LDPM constitutive laws as described earlier in this paper.

{\bf Hessian factorization and convergence criteria}: The tangent stiffness operator is evaluated by means of coloring-aided compressed forward-mode Automatic Differentiation. Moreover, the static implementation uses an~adaptive arc-length method, which was shown to obtain quadratic convergence for the vast majority of loading conditions tested~\parencite{jia2024efficient}. The convergence criterion uses $\ell_2$ norm of residual forces normalized by the norm of external forces, the tolerance is set to $10^{-7}$ but is reduced in slowly converging time steps up to $10^{-4}$.

\begin{table}[tb!]
\centering
\begin{tabular}{lllll}
\toprule
ID &  Software & Solver Type & Algorithm & Mass  \\
\midrule
AE&  ABAQUS& Explicit & Cent. Difference & Lumped\\
JA& JAX-LDPM& Explicit & Cent. Difference &  Lumped \\
CI&  Chrono&Implicit & HHT & Consistent \\
OA&  OAS&  Implicit & Generalized-$\alpha$ & Consistent \\
CA& CAST3M &Implicit & Static  & N/A \\
MP& FEMultiPhys& Implicit & HHT & Lumped \\
JU& Julia LDPM & Implicit & Static, Arc-Length\\
\bottomrule
\end{tabular}
\caption{Basic information for the adopted solvers, more information is available in Ref.~\parencite{LDPM_report}.}
\label{tab:Adopted-Software}
\end{table}

\begin{table}[tb!]
\centering
\begin{tabular}{lllll}
\toprule
ID &  Hessian & Hessian Factorization & Convergence Criteria & Conv. Tolerance  \\
\midrule
CI&  elastic & beg. of time step & increments & $10^{-4}$\\
OA&  elastic &  beg. of simul. & residuals, increments, energy & $10^{-4}$\\
CA&  elastic &  beg. of simul. & residuals & $10^{-4}$\\
MP&  elastic &  beg. of simul. & residuals & $10^{-3}$\\
JU&  tangent &  every iteration & residuals & $10^{-7}-10^{-4}$ \\
\bottomrule
\end{tabular}
\caption{Information about the Newton-Raphson iteration strategy in implicit solvers: stiffness matrix type, frequency of its evaluation and factorization, convergence criteria and thresholds.}
\label{tab:Implicit-Solvers}

\end{table}

\section{Numerical examples}

The various implementations are compared by simulating identical mechanical problems that include complex compressive and fracture behavior. All tests utilize the same material, with a~sieve curve defined by a~Fuller curve (exponent 0.4), maximum aggregate size of 12\,mm, cement content of $c=375$\,kg/m$^3$, and water-to-cement ratio of $w/c=0.5$. The material parameters are identical to all the numerical examples and are listed in Tab.~\ref{t:matparams} in the appendix. These parameters were adopted from identified values based on available experimental data in Ref.~\parencite{cusatis2011lattice2}. Report~\parencite{LDPM_report} describes the verification simulations done on a~single LDPM tetraherdon with imposed kinematics to ensure that all implementations provided the exactly identical response of a~single LDPM tetrahedron.

The results of individual software implementations are compared as follows.
\begin{itemize}
\item The \emph{global response} is visually compared by plotting the magnitude of the applied load versus displacement. For some of the  the tests, nominal stress and nominal strain calculated from load and displacement are plotted instead. 
\item The \emph{number of iterations} for the \emph{implicit} solvers is reported for each time step. The explicit solvers AE and JA do not iterate and therefore are not included. 
\item The \emph{kinetic energy} $W_{\text{kin}}$ is plotted as function of time. 
The static solvers CA and JU are not included since they do not provide kinetic energy.
\item The relative \emph{energy balance} is computed as $100\times|(W_{\text{ext}}-W_{\text{int}}-W_{\text{kin}})/W_{\text{ext}}|$ and is ploted as function of time. $W_{\text{ext}}$ and $W_{\text{int}}$ denote the total work of internal or external forces, respectively. The steady-state solvers CA and JU are not compared by this characteristic due to the missing kinetic energy.
\item The \emph{crack pattern} at the end of the simulations is compared by computing the total crack opening $w = \sqrt{w_N^2+w^2_M+w^2_L}$ at each facet of the LDPM tetrahedrons; $w_{\alpha}$ denotes the crack opening scalar in the local reference system, with $\alpha=N$ being the normal direction for which only positive inelastic strains give rise to crack opening $w_N$. Then, both (i) Pearson's correlation coefficients and (ii) Normalized Root Mean Square Errors (NRMSE) are reported for each pair of implementations. The NRMSE reads
\begin{align}
\mathrm{NRMSE}(a,b) = \frac{100}{\tilde w}\sqrt{\frac{1}{n}\sum_{i=1}^n \left[w^{(a)}_i-w^{(b)}_i\right]^2}
\end{align}
where symbols $a$ and $b$ refer to the pair of implementations~compared, $n$ is the number of facets in the simulation, which is 12 times the number of LDPM tetrahedra. $\tilde w$ is the normalization constant~taken as the maximum crack opening $\max_i w^{(OA)}$ in the OA simulation, and the constant 100 is included to have the error in percent units.  

There is no generally accepted level of correlation or NRMSE to detect agreeing data sets. In this paper four different internals are arbitrarily recognized and identified by shade of gray with growing intensity: (i) practically identical -- correlation $<0.999$, NRMSE $<1$\,\%, (ii) minor differences -- correlation $<0.9$, NRMSE $<5$\,\%, (iii) major differences -- correlation $<0.8$, NRMSE $<10$\,\%, and (iv) largely different.

\item The \emph{performance} of the implementations is assessed by measuring the actual computational time. All the \emph{implicit} solvers run on the same processor, {\ttfamily Intel(R) Xeon(R) Gold 6230}, using a~single computational thread. The single thread simulation allows at least some basic comparison of computational performance among the implementations. However, most of the software allow for parallelization of the process that unlocks significant reduction of computational time.  The two \emph{explicit} solvers are allowed to use more computational threads because their parallelization is a~natural and essential feature. AE is run on the same processor as the implicit solvers, using 24 computational threads. The runtime scale almost linearly with number of computational threads~\parencite{Ber19} and therefore some basic comparison with implicit solvers is possible by multiplying its runtime by a~factor between 12 and 24. The other explicit code, JA, is specifically designed for Graphical Processing Units, it therefore runs on {\ttfamily NVIDIA Quadro RTX 8000} using all its available performance. Therefore, its reported runtime cannot be compared directly with other implementations.
\end{itemize}

\subsection{Free vibrations in linear elastic regime}
\begin{figure}[tb!]
     \centering
     \includegraphics[width=\textwidth]{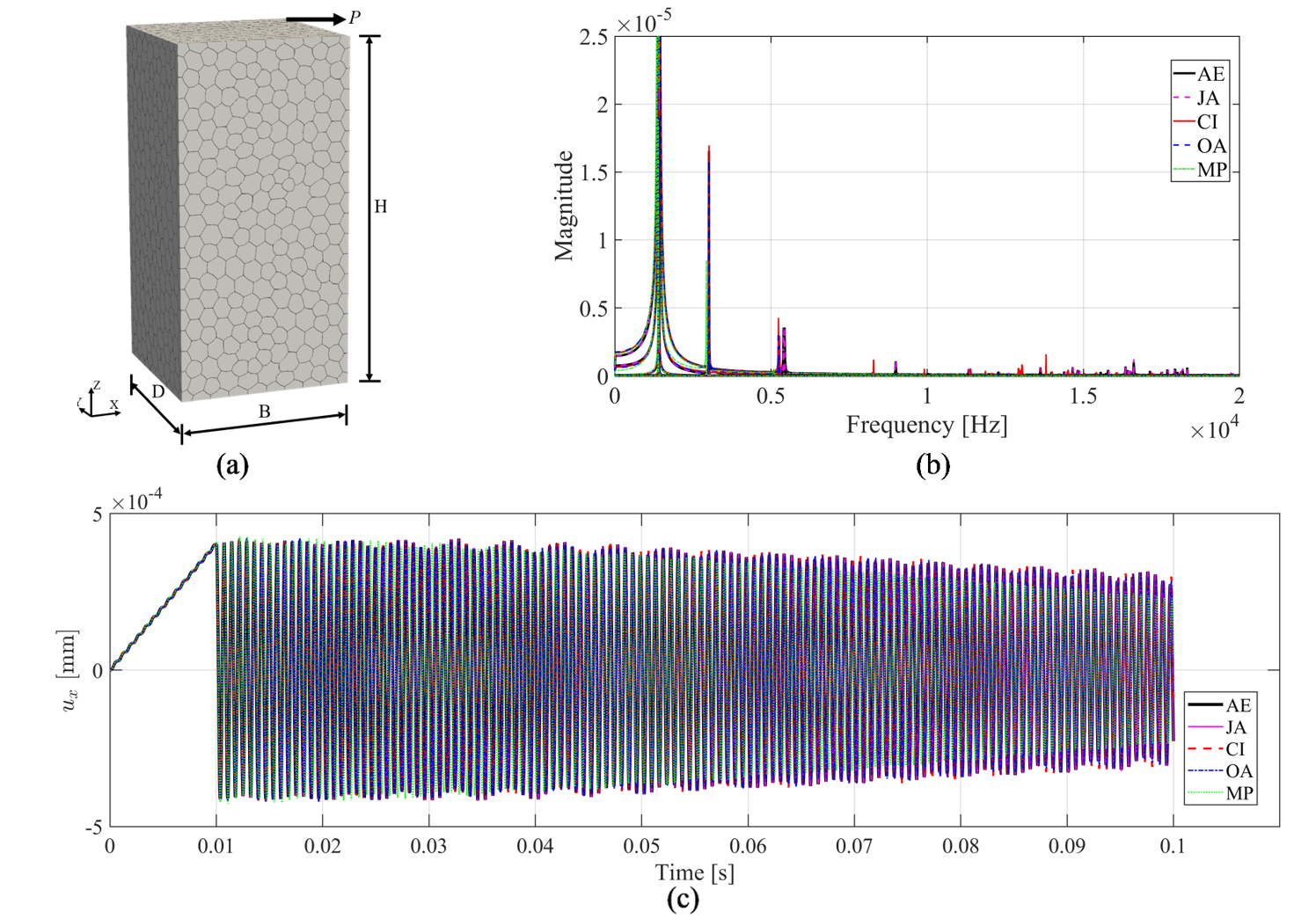}
     \caption{(a) dimensions of the specimen; (b) FFT of the response; (c) translations $u_x$ of the corner node in time.}
     \label{fig:FreeVibrations}
\end{figure}

The first benchmark simulates free vibrations of an~elastic prismatic sample ($D=B= 100$\,mm, $H = 200$\,mm) which is loaded by a~force $P$ applied at a~corner node located at the top end of the sample as shown in Figure~\ref{fig:FreeVibrations}a. Although all DoF at bottom nodes (including rotations) are restricted, top nodes are allowed to move freely. The force is applied in a~sudden manner by increasing its value linearly from $0$ to $50$\,N in a~very short time, $0.01$\,s, and deactivating the load linearly within another $0.00004$\,s. Then, the specimen is allowed to freely vibrate for a~total time of 0.1 sec.  The mechanical problem has approximately 15\,200\,DoF in total. In this benchmark,  linear elastic LDPM constitutive equations are used with the elastic parameters $E_0$ and $\alpha$ according to Tab.~\ref{t:matparams}. The steady-state (static) solvers were obviously omitted in this benchmark.

Figure~\ref{tab:Freqs}b reports the governing natural frequencies of the system obtained from the fast Fourier Transformation (FFT). The first few are also listed in Tab.~\ref{tab:Freqs}. Figure~\ref{fig:FreeVibrations}c shows $x$-displacement versus time. The first natural frequency $\approx$1400\,Hz corresponds to the large visible oscillations between positive and negative displacement values. The overall decrease of amplitude in time in Fig.~\ref{fig:FreeVibrations}c is not caused by damping but it is another low frequency oscillation that is not captured by FFT because the time series is too short.

In general, all the solvers provide practically identical response, the specific time integration method and implementation have almost no effect. The minor differences observed come from the representation of the mass matrix. The implicit solvers CI and OA with the consistent mass matrix report very similar results. 
These values are slightly different from those of the explicit solvers, which adopt a~lumped mass matrix (AE and JA) and suffer from well-known numerical dispersion \cite{mullen1982dispersion}. The frequency values calculated by the MP implicit solver differ from the others, albeit by just few percentage points, due to the approximation in calculating the moments of inertia of LDPM cells; see Sect.~\ref{sec:MP}.

\begin{table}[tb!]
    \centering
    \begin{tabular}{clllll}
       \toprule
       Frequency \# & AE & JA & CI & OA & MP \\ \midrule
       1 & 1404.79 & 1404.79  & 1404.23 & 1403.09 & 1381.19\\
       2 & 3009.09 & 3009.09  & 3009.54 & 3008.41 & 2928.21\\
       3 & 5399.68 & 5399.45  & 5234.04 & 5223.81 & 5069.29\\
       4 & 5435.92 & 5435.70  & 5265.85 & 5255.62 & 5105.63\\
       5 & 8991.48 & 8991.14  & 8281.07 & 8236.76 & -\\
       \bottomrule  
    \end{tabular}
    \caption{First five frequencies [Hz]  with peak amplitudes for \emph{free vibration} test in ascending order.}
    \label{tab:Freqs}
\end{table}

 Table~\ref{tab:FreeVibrations} reports the time steps used by different solvers: $10^{-7}$\,s for explicit solvers and $2\times10^{-5}$\,s for implicit solvers, and total computational times in hours. The explicit solvers are conditionally stable and the simulations were performed with a~time step approximately equal to 90~\% of the critical time step estimated from eigenvalue analysis at the level of individual tetrahedron~\parencite{cusatis2011lattice1,BelLiuMor00}. For this simulation the implicit solvers perform better than the explicit solvers since the system is linear and no iterations are required. This is true even if the explicit calculations are performed using several computational threads. 
Further speed-up of the implicit solvers can be gain by built-in parallelization.

\begin{table}[tb!]
    \centering
    \begin{tabular}{llllll}
       \toprule
        & AE & JA & CI & OA & MP\\ \midrule
       time step [s] & $10^{-7}$ & $10^{-7}$ & $2\times 10^{-5}$ & $2\times 10^{-5}$ & $2\times 10^{-5}$\\
       time [h] & 2.5 & 0.97  & 2.8 & 0.45 & 0.5\\
       \bottomrule  
    \end{tabular}
    \caption{Time step and total computational time for \emph{free vibration} test.}
    \label{tab:FreeVibrations}
\end{table}

\subsection{Hardening behavior in uniaxial strain compressive test}

\begin{figure}[tb!]
\centering
\includegraphics[width=\textwidth]{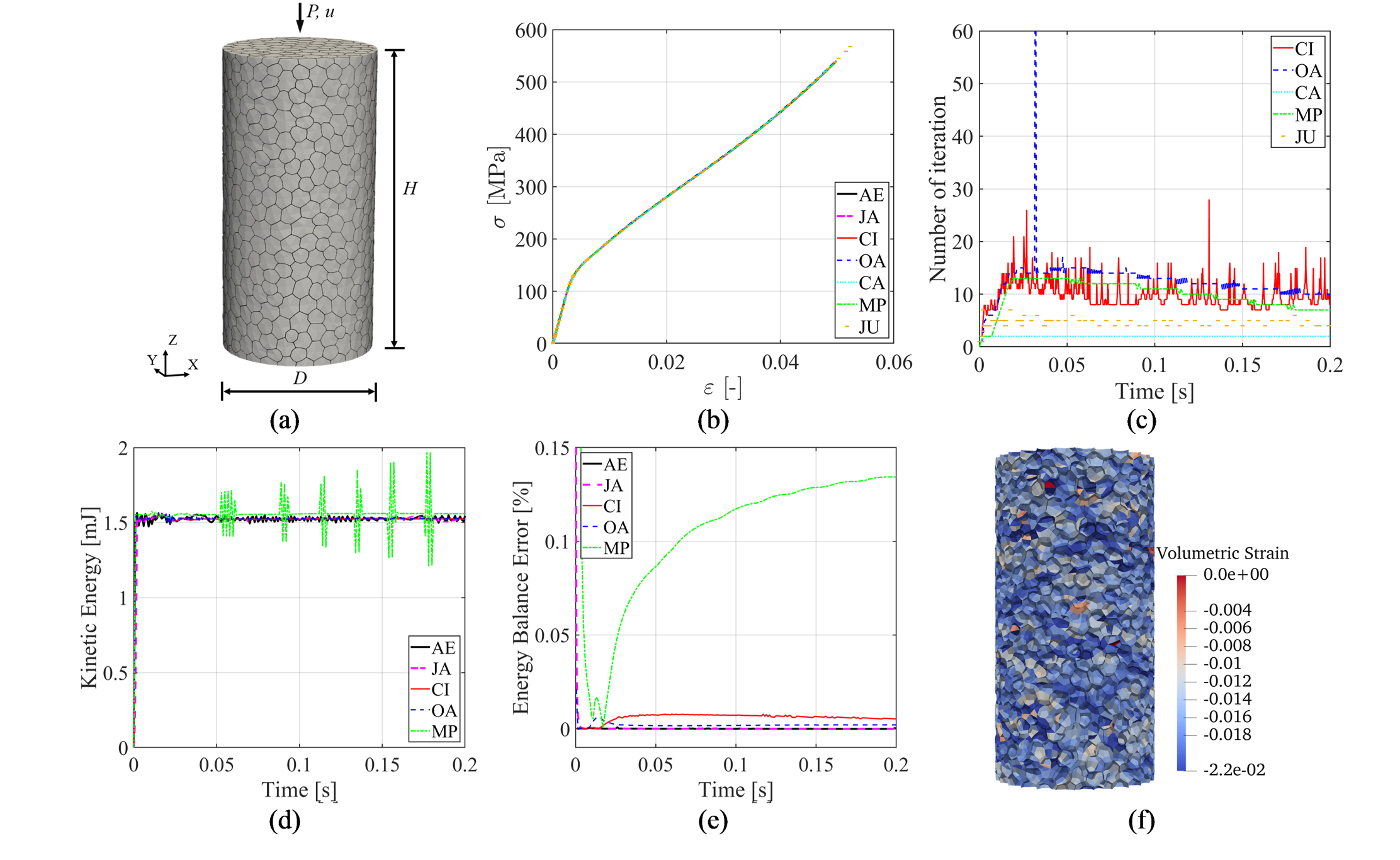}
\caption{The \emph{uniaxial strain} test: (a) dimensions of the specimen; (b) nominal stress-strain response in the vertical direction; (c) number of iterations of implicit solvers; (d) kinetic energy; (e) percentage error in energy balance; (f) distribution of the volumetric strain scalar in the final time step.}
\label{fig:Uniaxialmodel}
\end{figure}

The next simulated test is the~uniaxial strain test. The cylindrical sample ($D = 100$\,mm, $H= 200$\,mm) depicted in Figure~\ref{fig:Uniaxialmodel}a is subject to compression by applying velocity $\dot{u}_z=-5$\,mm/s to all nodes located at the top end of the sample. All nodes at the bottom of the specimen are constrained in the $z$-direction. Furthermore, the rotations $\theta_x$ and $\theta_y$ at both top and bottom ends of the specimens are restricted. Finally, lateral expansion is prevented by constraining the surface nodes in the $x$ and $y$ directions. The loading velocity is actually not constant, but grows linearly from zero to $-5$\,mm/s within the first 0.001 \,s to reduce abrupt changes in the loading process. The mechanical problem has approx.~10\,900\,DoF. 

Figure~\ref{fig:Uniaxialmodel}b reports the calculated nominal (macroscopic) stress (load force over area, $-4P/\pi D^2$) versus the imposed nominal (macroscopic) strain (top displacement over length, $-u_z/H$). All the different implementations provide virtually identical response featuring strain-hardening. The number of iterations for the implicit solvers is reported in Fig.~\ref{fig:Uniaxialmodel}c and is relatively stable except for one event for which the OA solver needed an~exceptionally large number of iterations. The reported kinetic energy (Fig.~\ref{fig:Uniaxialmodel}d) and the energy balance error (Fig.~\ref{fig:Uniaxialmodel}e) confirm that all simulations are stable and close to steady state  (terminal internal energy is 25\,kJ). This is also confirmed by including the steady-state simulations by CA and JU solvers, which are indistinguishable from the transient responses. The largest energy error and largest oscillations in kinetic energy are shown by MP solver and they ca be attributed by the approximated character of certain terms of the mass matrix. 
Finally, Fig.~\ref{fig:Uniaxialmodel}f shows the distribution of the volumetric strain (by CI solver) in the terminal step calculated from the tetrahedral LDPM elements. It is interesting to see the large spatial variability of the volumetric strain due to the inherent ability of LDPM to capture the effect of material heterogeneity even when the material is subject to a~macroscopically uniform state. 

The crack openings are not analyzed for this setup as they are not the most important outputs of the models. Instead, the volumetric strain computed at the end of the simulations is compared. Pearson's correlation coefficient and NRMSE of volumetric strains
for each pair of simulations are reported in Tab.~\ref{tab:UniaxialCrack}. Solvers AE, JA, CI and CA provides virtually identical results as the correlations are 1 and NRMSE is below 0.2\,\%. Results from OA, MP and JU deviate a~little but the differences are still small.

\begin{table}[tb!]
\begin{center}
\begin{tabular}{c|ccccccc}
 & AE & JA & CI & OA & MP & CA & JU\\\hline
AE&&0.050 &0.139 &\cellcolor{gray!40} 2.014 &\cellcolor{gray!40} 1.157 &0.187 &\cellcolor{gray!40} 1.479 \\
JA&1.000 &&0.142 &\cellcolor{gray!40} 2.009 &\cellcolor{gray!40} 1.159 &0.178 &\cellcolor{gray!40} 1.510 \\
CI&1.000 &1.000 &&\cellcolor{gray!40} 2.013 &\cellcolor{gray!40} 1.169 &0.104 &\cellcolor{gray!40} 1.454 \\
OA&\cellcolor{gray!40} 0.935 &\cellcolor{gray!40} 0.935 &\cellcolor{gray!40} 0.935 &&\cellcolor{gray!40} 2.376 &\cellcolor{gray!40} 2.025 &\cellcolor{gray!40} 2.423 \\
MP&\cellcolor{gray!40} 0.979 &\cellcolor{gray!40} 0.979 &\cellcolor{gray!40} 0.979 &\cellcolor{gray!40} 0.910 &&\cellcolor{gray!40} 1.173 &\cellcolor{gray!40} 1.945 \\
CA&0.999 &0.999 &1.000 &\cellcolor{gray!40} 0.934 &\cellcolor{gray!40} 0.979 &&\cellcolor{gray!40} 1.502 \\
JU&\cellcolor{gray!40} 0.985 &\cellcolor{gray!40} 0.985 &\cellcolor{gray!40} 0.986 &\cellcolor{gray!40} 0.934 &\cellcolor{gray!40} 0.960 &\cellcolor{gray!40} 0.986 &\end{tabular}
\end{center}
\caption{Correlations (lower triangle) and NRMSE (upper triangle) of volumetric strain at the end ($t=0.2$\,s) of \emph{uniaxial strain} simulation. \label{tab:UniaxialCrack}}
\end{table}

\begin{table}[tb!]
    \centering
    \begin{tabular}{llllllll}
       \toprule
       & AE & JA & CI & OA & MP  & CA & JU\\ \midrule
       time step [s] & $10^{-7}$ & $10^{-7}$ & $ 10^{-4}$ & $5\times 10^{-4}$ & $2\times 10^{-5}$ & $10^{-4}$ & -\\
       time [h] & 2.2 & 1.6 & 1.9 & 0.25 & 0.22 & 8.3 & 6.6\\
       \bottomrule  
    \end{tabular}
    \caption{Time step and total computational time for \emph{uniaxial strain} test.}
    \label{tab:Uniaxialmodel}
\end{table}

The time steps and computational times are reported in Tab.~\ref{tab:Uniaxialmodel}. Due to the relative simplicity of this hardening example and the consequently low number of iterations, the implicit solvers are often advantageous, even though they all run on a~single computational thread, as they can use much larger time steps. The CA and JU are slower most likely because they update and factorize the stiffness matrix each time step (CA, elastic matrix) or every iteration (JU, tangent matrix). However, the extra work does not bring real speed-up since the problem seems to converge fast even with the elastic stiffness matrix.

\subsection{Fracturing behavior in dog bone tensile test}

\begin{figure}[tb!]
\centering
\includegraphics[width=\textwidth]{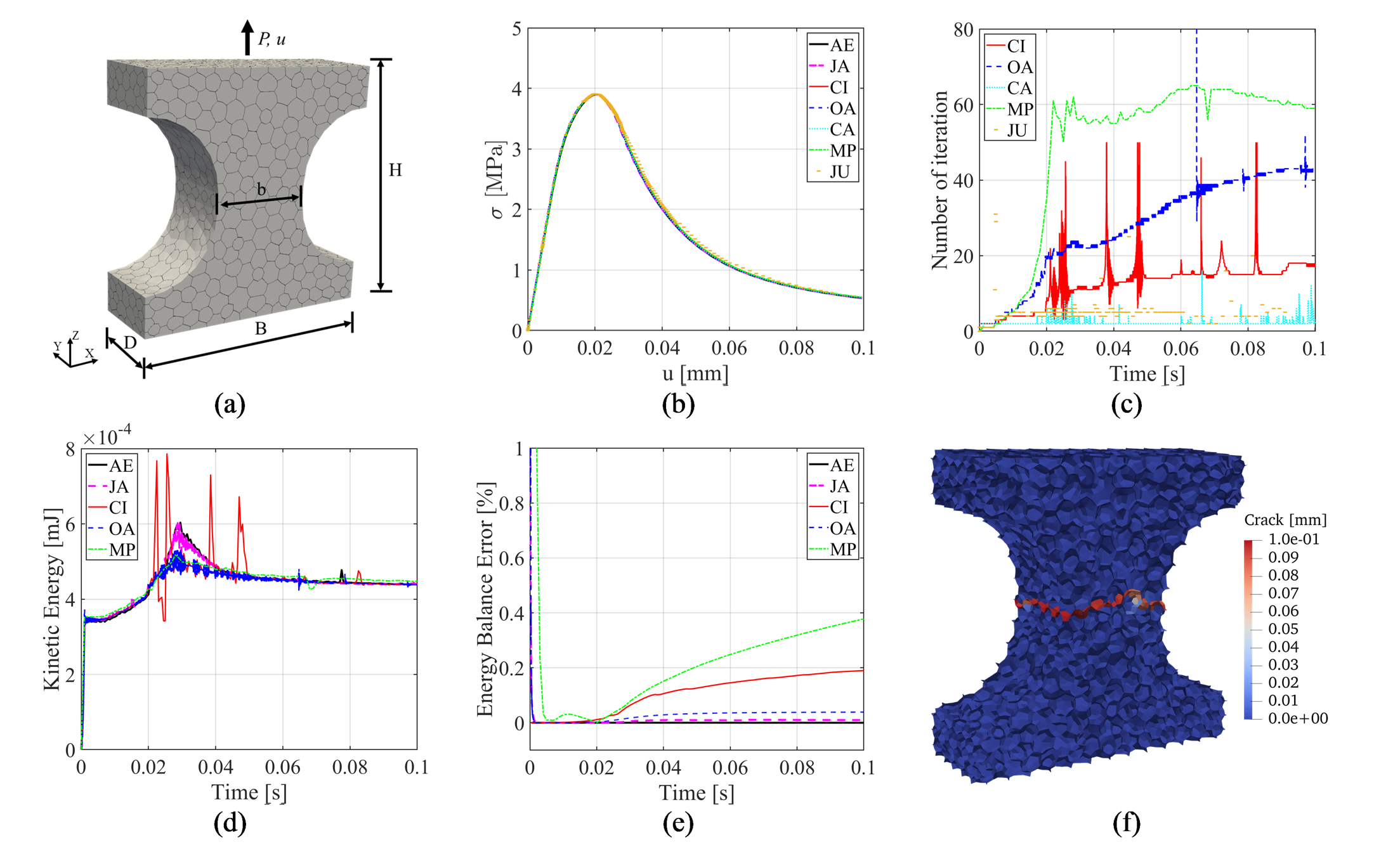}
\caption{The \emph{dog bone} test: (a) dimensions of the specimen; (b) stress-displacement response in the vertical direction; (c) number of iterations of implicit solvers; (d) kinetic energy; (e) percentage error in energy balance; (f) crack pattern in the final time step.}
\label{fig:dogbone}
\end{figure}

The third benchmark features a~dog bone shaped specimen loaded in tension. The specimen is sketched in Fig.~\ref{fig:dogbone}a, the dimensions are $B=H=150$\,mm, $b=D=50$\,mm. All translations and rotations $\theta_x$ and $\theta_y$ of all lower surface nodes are restricted, the same applies to the upper surface nodes, except the vertical velocity is prescribed, $\dot{u}_z = 1$\,mm/s. Similarly to the uniaxial compression test, the velocity increases linearly from zero within first 0.001\,s. The loading force $P$ is measured. There are approximately 6\,600 DoF in this simulation.

The stress-displacement data obtained are reported in Fig.~\ref{fig:dogbone}b, stress is related to the smallest central cross section, $\sigma=P/bD$. All implementations provide almost  indistinguishable responses, including the static solvers. The number of iterations in each time step for implicit solvers is given in Fig.~\ref{fig:dogbone}c, again one step in the OA solver did not converge. The JU solver, which predicts slightly higher forces than the others,  did not finish the calculation and stopped at 0.06\,s due to convergence problems. The kinetic energy and energy error are shown in Figs.~\ref{fig:dogbone}d and \ref{fig:dogbone}e. The kinetic energy is relatively small (the internal energy in the terminal step is about 500\,mJ), and the energy error does not show any unstable behavior. Fig.~\ref{fig:dogbone}f shows the crack pattern at the end of the simulation (by CI), the colors indicate the total crack opening computed $w$. The response is characterized by only one localized fracture developed in the specimen.

\begin{table}[tb!]
\begin{center}\begin{tabular}{c|ccccccc}
 & AE & JA & CI & OA & MP & CA & JU\\\hline
AE&&0.258 &0.467 &0.479 &0.436 & \cellcolor{gray!70} 7.778 &0.533 \\
JA&1.000 &&0.360 &0.404 &0.278 & \cellcolor{gray!70} 7.774 &0.399 \\
CI&0.999 &1.000 &&0.249 &0.341 & \cellcolor{gray!70} 7.764 &0.109 \\
OA&0.999 &1.000 &1.000 &&0.433 & \cellcolor{gray!70} 7.767 &0.310 \\
MP&0.999 &1.000 &1.000 &0.999 && \cellcolor{gray!70} 7.729 &0.345 \\
CA& \cellcolor{gray!70} 0.859 & \cellcolor{gray!70} 0.860 & \cellcolor{gray!70} 0.860 & \cellcolor{gray!70} 0.860 & \cellcolor{gray!70} 0.861 && \cellcolor{gray!70} 7.767 \\
JU&0.999 &1.000 &1.000 &1.000 &1.000 & \cellcolor{gray!70} 0.860 &\end{tabular}
\end{center}
\caption{Correlations (lower triangle) and NRMSE (upper triangle) of crack openings at the end ($t=0.1$\,s) of \emph{dog bone} simulation. \label{tab:dogboneCracks}}
\end{table}

The comparison of the final crack patterns through correlation and NRMSE is presented in Tab.~\ref{tab:dogboneCracks}. Except CA, the numbers show an~excellent correspondence among all the solvers. Table~\ref{tab:dogbone} provides the time steps and total computational times for individual implementations. MP solver is the fastest this time, outperforming the explicit solvers using multiple computational threads. The fast MP computations are most likely achieved by using the largest time step and weakest convergence tolerance, on the other hand the number of iterations is typically around 60 in every time step which does not comply with the requirement to keep their number bellow 50. Other implicit solvers delivered the results in a few hours, which is still a~reasonable time considering that only a~single computational thread was utilized.

\begin{table}[tb!]
    \centering
    \begin{tabular}{llllllll}
       \toprule
        & AE & JA & CI & OA & MP & CA & JU \\ \midrule
       time step [s] & $10^{-7}$ & $10^{-7}$ & $2\times 10^{-5}$ & $2.5\times 10^{-5}$ & $5\times 10^{-4}$  & $10^{-4}$ & -\\
       time [h] & 1.33 & 0.43 & 3.9 & 3.2 & 0.14 & 1.1 & 11 \\
       \bottomrule  
    \end{tabular}
    \caption{Time step and total computational time for \emph{dog bone} tensile test.}
    \label{tab:dogbone}
\end{table}

\subsection{Fracturing in three-point bending test}

\begin{figure}[tb!]
\centering
\includegraphics[width=\textwidth]{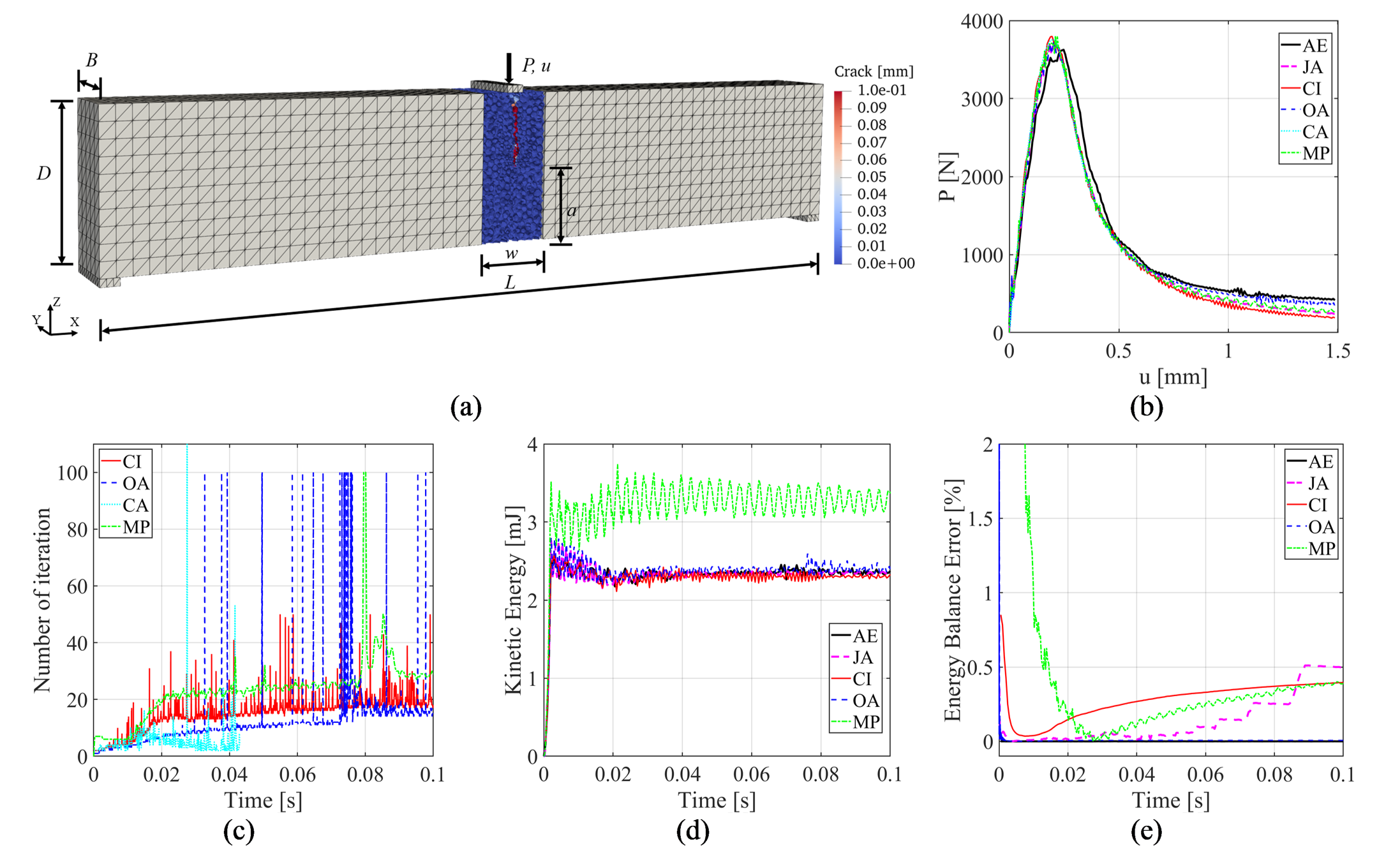}
\caption{The \emph{three-point bending} test: (a) dimensions of the specimen \& crack pattern in the final time step; (b) load-displacement response; (c) number of iterations of implicit solvers; (d) kinetic energy; (e) percentage error in energy balance.}
\label{fig:tpb}
\end{figure}

The classical three-point bending test with a~central notch is simulated. The specimen is sketched in Fig.~\ref{fig:tpb}a, the dimensions are $L=1151$\,mm, $D=200$\,mm, $B=100$\,mm, notch depth is $a=100$\,mm (50\,\% of $D$), notch width is 4\,mm. For the sake of computational time reduction, only the central part of width $w=100$\,mm is represented by the mesoscale discrete model, the rest of the material is modeled by a~continuous homogeneous linear elastic model (standard tri-linear tetrahedral finite elements) with elastic modulus 39\,GPa, Poisson's ratio 0.2 and density 2380\,kg/m$^3$. The platens are treated as rigid bodies with a~density of 7800\,kg/m$^3$. The connection between the continuous and the discrete part is provided by a~linear constraint, and the boundary particle translations are governed by the shape functions of the adjacent tetrahedral elements. This dependency is realized in each software differently, either by direct reduction of the degrees of freedom or by Lagrangian multipliers. The boundary condition restricts $u_z$ and $u_y$ movements of the central ($y$-direction) bottom ($z$-direction) most left and right ($x$-direction) nodes of the concrete part and both horizontal movements of the central top node. The vertical displacement of the central top node is prescribed with velocity $-15$\,mm/s with the initial linear increase period of duration 0.002\,s. The number of DoF is approximately 27\,000.

The loading force and Crack Mouth Opening Displacement (CMOD) are reported in Fig.~\ref{fig:tpb}b. This time, there are visible differences in the curves, although the overall character of the response is the same. The differences are attributed to the different treatment of the connection between the continuous and discrete domains, as it appears already in early stages of the simulation.  The number of iterations in the implicit solvers is shown in Fig.~\ref{fig:tpb}c and the kinetic energy and the energy balance error are reported in Figs.~\ref{fig:tpb}c and d, respectively. Most of the implicit solvers suffer from poor convergence, and the non-converged solution is accepted several times. The CA solver even terminated soon after the peak force because of divergent iterations. The results of the JU solver are not reported as the solver convergence was lost too soon. Moreover, one can see relatively large energy balance errors for CI, JA and MP. In addition, MP provides significantly higher kinetic energy, which is probably attributed to different treatment of the mass matrix. Nevertheless, the difference in the computed force is relatively small.

\begin{table}[tb!]
\begin{center}\begin{tabular}{c|ccccc}
 & AE & JA & CI & OA & MP\\\hline
AE&&\cellcolor{gray!40} 3.428 &\cellcolor{gray!40} 3.467 &\cellcolor{gray!40} 2.958 &\cellcolor{gray!40} 3.000 \\
JA&\cellcolor{gray!40} 0.944 &&0.442 &\cellcolor{gray!40} 1.752 &\cellcolor{gray!40} 1.758 \\
CI&\cellcolor{gray!40} 0.944 &0.999 &&\cellcolor{gray!40} 1.828 &\cellcolor{gray!40} 1.786 \\
OA&\cellcolor{gray!40} 0.958 &\cellcolor{gray!40} 0.985 &\cellcolor{gray!40} 0.984 &&0.355 \\
MP&\cellcolor{gray!40} 0.958 &\cellcolor{gray!40} 0.986 &\cellcolor{gray!40} 0.985 &1.000 &\end{tabular}
\end{center}
\caption{Correlations (lower triangle) and NRMSE (upper triangle) of crack openings at the end ($t=0.1$\,s) of \emph{three-point bending} simulation.\label{tab:tpbCracks}}
\end{table}

Contrary to surprising variations in load force, the comparison of crack patterns show only mild differences. The NRMSE of crack opening in Tab.~\ref{tab:tpbCracks} is about 2-3\,\% for most of the pairs under comparison, the correlation remains close to one. By visually comparing the cracks, the differences occur predominantly in the area right below the loading platen. Some solvers predict the final crack growth towards the left side of the platen (AE,OA,MP), while some show cracking at both sides (CI,JA). Note that pairs OA--MP and CI--JA predicts practically identical cracks. The AE solver exhibit the most different crack and also the most different loading force.

Table~\ref{tab:tpb} provides the time steps and computational times for this benchmark with quite a~large scatter. CI and OA needed the longest time, partly due to the small time step and partly due to strict convergence criteria $10^{-4}$. CA time is not comparable as it did not finish the whole simulation. MP with the weakest converge threshold reports again the fastest results from all the implicit solvers. The explicit solvers are performing well and for the first time they might be considered advantageous over most of the implicit solvers.

\begin{table}[tb!]
    \centering
    \begin{tabular}{lllllll}
       \toprule
        & AE & JA & CI & OA & MP & CA\\ \midrule
       time step [s] & $10^{-7}$ & $10^{-7}$ & $10^{-5}$ & $2\times 10^{-6}$ & $2\times 10^{-5}$ & $10^{-4}$\\
       time [h] & 3.7 & 1.27 & 24.5 & 20.8 & 5.0 & 1.6$^{\star}$\\
       \bottomrule  
    \end{tabular}
    \caption{Time step and total computational time for notched \emph{three-point bending} test.  Symbol $^{\star}$ indicates simulations terminated at loading time 0.43\,s before reaching the final time step.}
    \label{tab:tpb}
\end{table}

\subsection{Unconfined compression test with pervasive fracturing}

\begin{figure}[tb!]
\centering
\includegraphics[width=\textwidth]{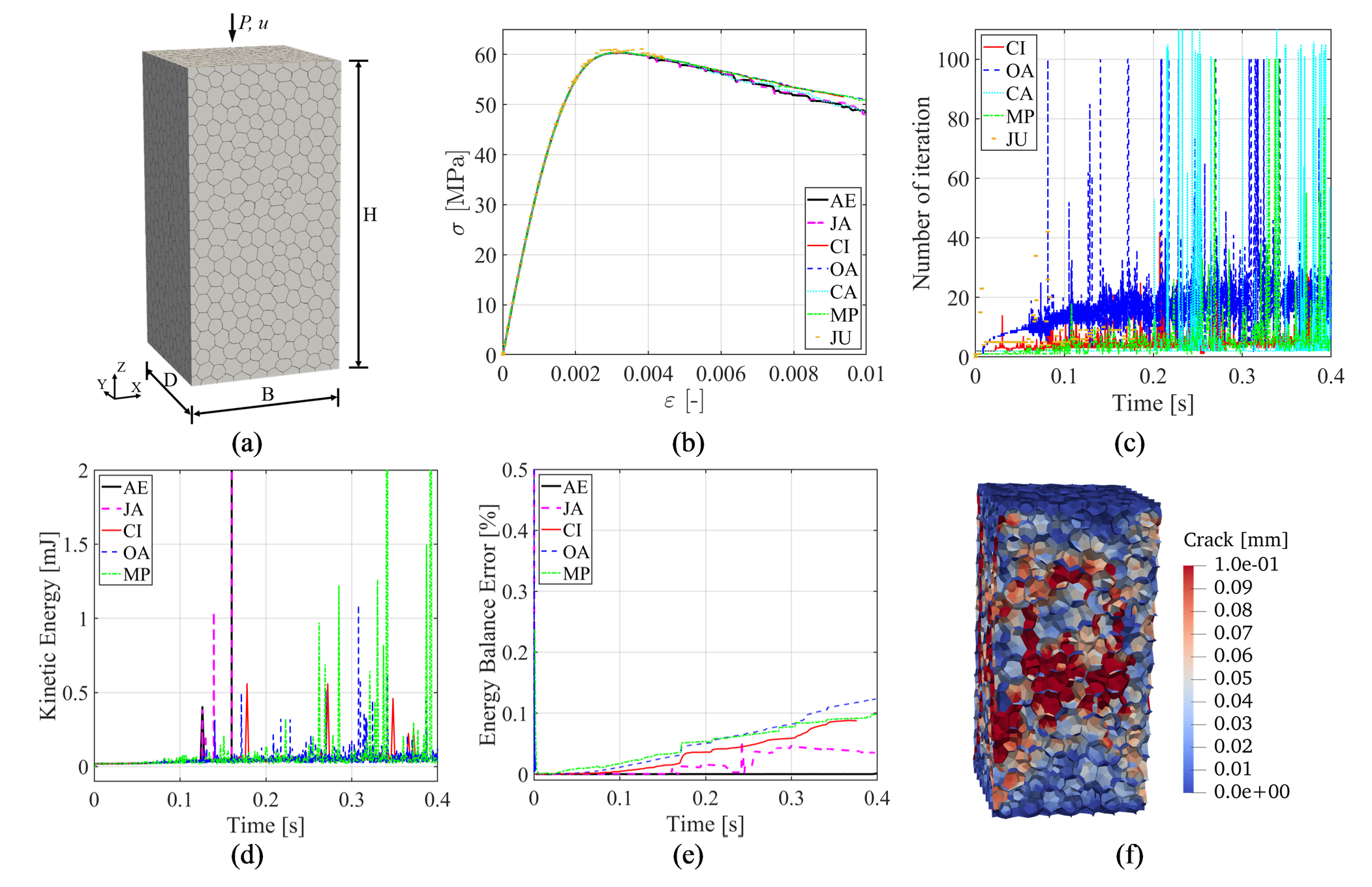}
\caption{The \emph{fixed unconfined compression} test: (a) dimensions of the specimen; (b) stress-strain response in the vertical direction; (c) number of iterations of implicit solvers; (d) kinetic energy; (e) percentage error in energy balance; (f) crack pattern in the final time step.}  
\label{fig:ucfix}
\end{figure}

In this section, an~unconfined compression test is simulated on a~prismatic sample sketched in Fig.~\ref{fig:ucfix}a, $B=D=100$\,mm, $H=200$\,mm. The friction forces between the loading platens at the top and bottom surfaces of the specimen have a~large influence on the mechanical response, as well as on the spatial crack distribution~\parencite{cusatis2011lattice2}. In order to account for different typical situations, two extreme cases are considered. In the first case, it is assumed that the friction is large enough to completely eliminate horizontal movements on the top and bottom surfaces. This high-friction case is hereinafter referred to as \emph{fixed}. For the second extreme case, friction is negligible, and the top and bottom surface nodes can freely move horizontally. This scenario is referred to as \emph{free} in this study.

In the fixed case, all nodes on the bottom surface have all displacements and rotations restricted. The same is valid for the top surface, except that the vertical velocity $\dot{u}_z=-5$\,mm/s is applied (after the initial linear ramp of duration 0.002\,s). In the free case, the same applies, but the horizontal displacements $u_x$ and $u_y$ and the rotation $\theta_z$ are left free on both surfaces. Moreover, the central nodes on the top and bottom surfaces have both horizontal displacements and rotation $\theta_z$ restricted. There is approx. 14\,000 DoF in the model of both scenarios.

The results of the fixed simulations are reported in Fig.~\ref{fig:ucfix}. All implementations give almost identical results in terms of the stress-strain response until the strain reaches approx.~$0.004$. The agreement among the different response curves deteriorates to some extent after the peak in the descending portion of the curve. In this regime, the response is characterized by spikes in kinetic energy (Fig. \ref{fig:ucfix}) associated with the dynamic detachment of material fragments from the surface of the sample, especially along the vertical edges. This post-peak dynamic phenomenon is also observed in experiments regardless of the rate at which the loading plate displacement is applied. In correspondence of these dynamic events, the implicit solvers do not occasionally converge even for a~large number of iterations ($>100$). However, convergence is regained in subsequent steps when the spike of kinetic energy is dissipated due to material failure and nonlinear behavior of the material.  The explicit solvers AE and JA as well as the steady-state solver CA deviate from the others. Repetitive runs with different solution parameters result in an~identical discrepancy. The JU implementation is terminated earlier for loss of convergence.  

\begin{figure}[tb!]
\centering
\includegraphics[width=\textwidth]{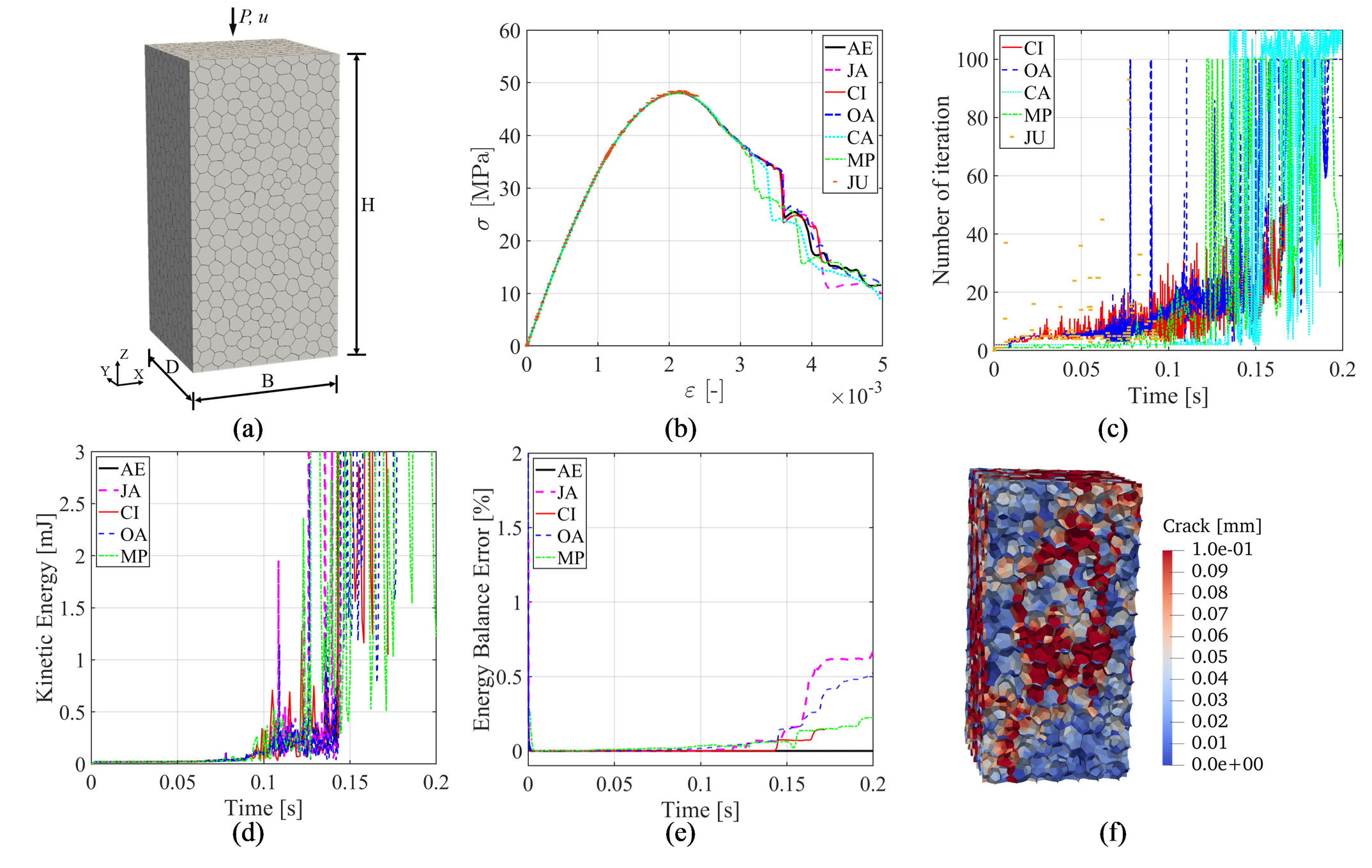}
\caption{The \emph{free unconfined compression} test: (a) dimensions of the specimen; (b) stress-strain response in the vertical direction; (c) number of iterations of implicit solvers; (d) kinetic energy; (e) percentage error in energy balance; (f) crack pattern in the final time step.}  
\label{fig:ucfree}
\end{figure}

Figure~\ref{fig:ucfree} presents the results obtained for the free case. All solvers give identical results up to a~macroscopic strain of approximately $3.5\times10^{-3}$ (time 0.14\,s). The response up to the maximum load is basically quasi-static as shown by the almost negligible kinetic energy (Fig. \ref{fig:ucfree}) for $t<0.1$\,s. After the peak load, the macroscopic stress initially decreases gradually with increasing deformation. In this phase, similarly to the previously discussed ``fixed'' case, the response starts to evolve dynamically, and the implicit solvers have instances in which they do not converge. 

After this initial softening phase, the response exhibits an~abrupt dynamic evolution of multiple fractures and is associated with a~sudden loss of macroscopic load carrying capacity (Fig. \ref{fig:ucfree}b). At this time, the specimen experiences pervasive fragmentation with spalling of fragments at high kinetic energy. The overall kinetic energy of the system increases significantly (Fig. \ref{fig:ucfix} for $t>$0.14\,s). All implicit solvers do not achieve convergence for a~large number of steps, and the accuracy of the solution is necessarily compromised. In addition, the explicit AE and JA solvers also provide quite different answers in this simulation phase. That means that from time $\approx0.14$\,s none of the tested solvers can deliver reliable results. 

The crack patterns obtained are given in Figs.~\ref{fig:ucfix}f and \ref{fig:ucfree}f for the fixed and free case, respectively. Both  figures are in the final step of CI implementation. As expected, the crack pattern occurs in a~pervasive manner in both cases. In addition, in the fixed case, cracks are not present at the top and bottom boundaries due to lateral constraints. The pair comparison by correlation and NRMSE is given in Tabs.~\ref{tab:ucfixedCrack} and \ref{tab:ucfreeCrackEnd}. Both tables are calculated on modified data where the maximum crack opening was limited to 4\,mm. This is done to eliminate the influence of detached fragments on the comparison of crack patterns. In both free and fixed scenarios, one can see large differences among the implementations. The only similarity is seen in the fixed case for CI, OA, and MP solvers; these solvers also provide identical stress response in the fixed case. A~visual comparison of the cracks reveals that they are still quite similar even when their correlations and NRMSE testify otherwise. The large number of fractured facets throughout the entire volume of the specimen leads to the accumulation of smaller differences into a~large discrepancy in the global characteristics. Note that a~large difference is seen also for the pair AE--JA of explicit solvers that are usually considered robust and reliable.

Additionally, a~crack comparison was also performed for the free case at time 0.12\,s, i.e., before the abrupt rupture of the specimen. Table~\ref{tab:ucfreeCrackMid} documents that, apart from that CA implementation, the differences are reasonably small.

The time steps and computational times are provided in Tab.~\ref{tab:ucfreefix}. The simulation times of implicit solvers using a~single thread are often tens of hours, especially for the free case. The explicit solvers, which utilize many computational threads, are typically faster in these simulations.

\begin{table}[tb!]
\begin{center}
\begin{tabular}{c|cccccc}
 & AE & JA & CI & OA & MP & CA\\\hline
AE&& \cellcolor{gray!70} 7.668 & \cellcolor{gray!70} 9.056 & \cellcolor{gray!70} 9.193 & \cellcolor{gray!70} 8.922 & \cellcolor{gray!70} 9.728 \\
JA& \cellcolor{gray!70} 0.828 && \cellcolor{gray} 11.540 & \cellcolor{gray} 11.681 & \cellcolor{gray} 11.453 & \cellcolor{gray} 10.866 \\
CI& \cellcolor{gray} 0.628 & \cellcolor{gray} 0.547 &&0.793 &\cellcolor{gray!40} 1.569 & \cellcolor{gray!70} 7.624 \\
OA& \cellcolor{gray} 0.615 & \cellcolor{gray} 0.534 &\cellcolor{gray!40} 0.993 &&\cellcolor{gray!40} 1.627 & \cellcolor{gray!70} 7.817 \\
MP& \cellcolor{gray} 0.648 & \cellcolor{gray} 0.565 &\cellcolor{gray!40} 0.972 &\cellcolor{gray!40} 0.968 && \cellcolor{gray!70} 7.853 \\
CA& \cellcolor{gray} 0.601 & \cellcolor{gray} 0.612 & \cellcolor{gray} 0.668 & \cellcolor{gray} 0.652 & \cellcolor{gray} 0.653 &\end{tabular}
\end{center}
\caption{Correlations (lower triangle) and NRMSE (upper triangle) of crack openings at the end ($t=0.4$\,s) of \emph{unconfined compressions} simulation with high friction (fixed).\label{tab:ucfixedCrack}}
\end{table}

\begin{table}[tb!]
\begin{center}
\begin{tabular}{c|cccccc}
 & AE & JA & CI & OA & MP & CA\\\hline
AE&& \cellcolor{gray} 13.458 & \cellcolor{gray!70} 8.332 & \cellcolor{gray!70} 6.081 & \cellcolor{gray!70} 8.315 & \cellcolor{gray} 14.023 \\
JA& \cellcolor{gray} 0.683 && \cellcolor{gray} 12.496 & \cellcolor{gray} 13.615 & \cellcolor{gray} 15.160 & \cellcolor{gray} 15.745 \\
CI& \cellcolor{gray} 0.798 & \cellcolor{gray} 0.784 && \cellcolor{gray!70} 6.327 & \cellcolor{gray!70} 9.918 & \cellcolor{gray} 15.314 \\
OA& \cellcolor{gray!70} 0.898 & \cellcolor{gray} 0.686 & \cellcolor{gray!70} 0.847 && \cellcolor{gray!70} 8.257 & \cellcolor{gray} 14.329 \\
MP& \cellcolor{gray!70} 0.820 & \cellcolor{gray} 0.577 & \cellcolor{gray} 0.719 & \cellcolor{gray!70} 0.817 && \cellcolor{gray} 15.081 \\
CA& \cellcolor{gray} 0.710 & \cellcolor{gray} 0.624 & \cellcolor{gray} 0.671 & \cellcolor{gray} 0.720 & \cellcolor{gray} 0.631 &\end{tabular}
\end{center}
\caption{Correlations (lower triangle) and NRMSE (upper triangle) of crack openings at the end ($t=0.2$\,s) of \emph{unconfined compressions} simulation  with low friction (free).\label{tab:ucfreeCrackEnd}}
\end{table}

\begin{table}[tb!]
\begin{center}\begin{tabular}{c|cccccc}
 & AE & JA & CI & OA & MP & CA\\\hline
AE&&0.054 &\cellcolor{gray!40} 4.828 &\cellcolor{gray!40} 4.769 &\cellcolor{gray!40} 4.473 & \cellcolor{gray} 10.087 \\
JA&1.000 &&\cellcolor{gray!40} 4.833 &\cellcolor{gray!40} 4.774 &\cellcolor{gray!40} 4.480 & \cellcolor{gray} 10.090 \\
CI&\cellcolor{gray!40} 0.938 &\cellcolor{gray!40} 0.938 &&\cellcolor{gray!40} 1.266 &\cellcolor{gray!40} 4.622 & \cellcolor{gray!70} 8.422 \\
OA&\cellcolor{gray!40} 0.936 &\cellcolor{gray!40} 0.936 &\cellcolor{gray!40} 0.995 &&\cellcolor{gray!40} 4.554 & \cellcolor{gray!70} 8.548 \\
MP&\cellcolor{gray!40} 0.946 &\cellcolor{gray!40} 0.946 &\cellcolor{gray!40} 0.930 &\cellcolor{gray!40} 0.931 && \cellcolor{gray!70} 8.748 \\
CA& \cellcolor{gray} 0.750 & \cellcolor{gray} 0.750 & \cellcolor{gray!70} 0.800 & \cellcolor{gray} 0.798 & \cellcolor{gray} 0.771 &\end{tabular}
\end{center}
\caption{Correlations (lower triangle) and NRMSE (upper triangle) of crack openings at time $t=0.12$\,s of \emph{unconfined compressions} simulation with low friction (free).\label{tab:ucfreeCrackMid}}
\end{table}

\begin{table}[tb!]
    \centering
    \begin{tabular}{lllllllll}
       \toprule
       &  & AE & JA & CI & OA & MP & CA & JU\\ \midrule
       \multirow{2}{*}{fixed} & time step [s] & $10^{-7}$ & $10^{-7}$ & $5\times 10^{-6}$ & $4\times 10^{-5}$ & $10^{-5}$  & $10^{-4}$ & - \\
       & time [h] & 4 & 5.3 & 62 & 4.6 & 7.5 & 18.5  & 5.4$^{\star}$\\
        \midrule       
       \multirow{2}{*}{free}& time step [s] & $10^{-7}$ & $10^{-7}$ & $5\times 10^{-6}$ & $2\times 10^{-5}$ & $10^{-5}$ & $10^{-4}$ & - \\
       & time [h] & 3.5 & 2.7 & 78.5 & 61 & 50 & 9.9 & 6.2$^{\star}$ \\
       \bottomrule  
    \end{tabular}
    \caption{Time steps and total computational times for \emph{unconfined compression} test with high and low friction. Symbol ${\star}$ indicates simulations terminated before reaching the end of loading due to convergence issues at times 0.18\,s (fixed) and 0.1\,s (free).}
    \label{tab:ucfreefix}
\end{table}

\section{Numerical study for time step, convergence criteria and stability}

As shown in the previous section, the most challenging simulation is the unconfined compression test without friction (free). This is due to a~transition from a~quasi-static behavior before the peak load to a~dynamic response in the post-peak phase, when fracture localization and fragmentation occur. 

\begin{figure}[tb!]
     \centering
     \includegraphics[width=15cm]{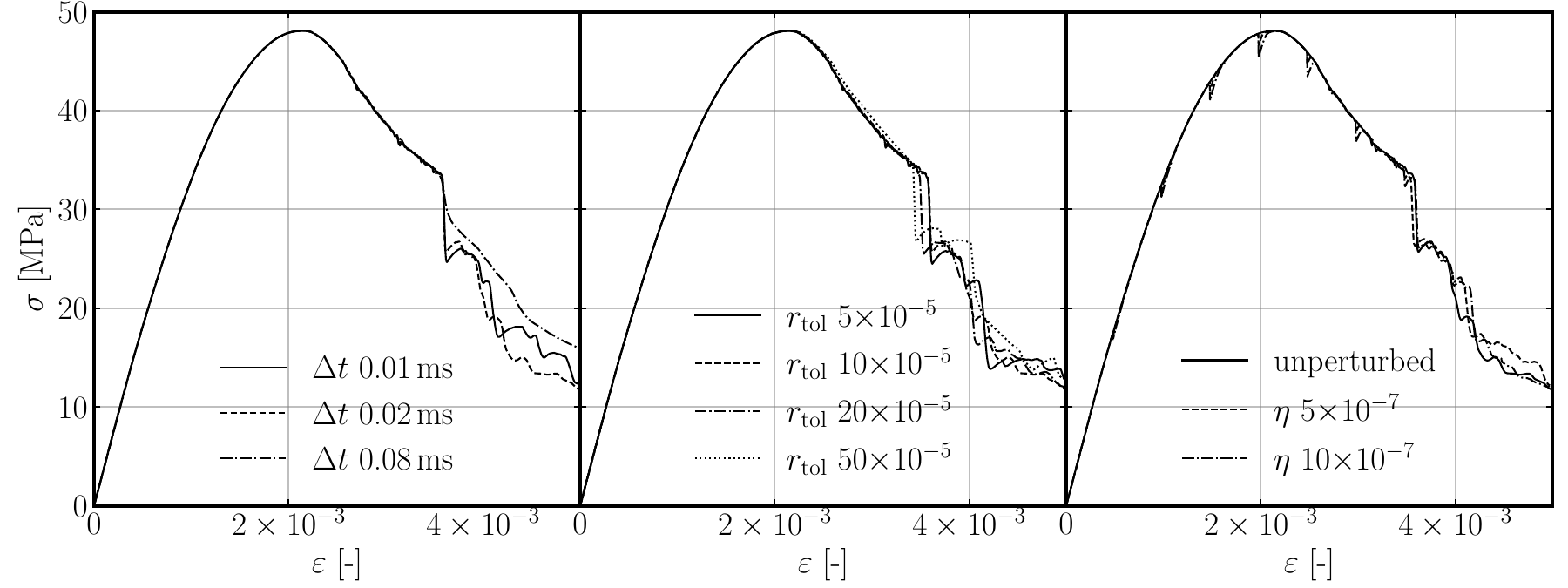}
     \caption{Sensitivity of the response of the unconfined compression test without friction to the time step length, convergence tolerance, and random perturbation.}
     \label{fig:ucfree_study}
\end{figure}

To analyze this case in detail, the OA solver is executed with (i) different time steps, (ii) different convergence tolerances, and (iii) periodic inserting of random perturbations into the solution. The results of these simulations are presented in Fig.~\ref{fig:ucfree_study}. The perturbations are introduced every 0.002\,s as a~random vector with a~uniform distribution ranging from $-\eta/2$ to $\eta/2$ added to  all free  displacements and rotations of all nodes. The remaining solver parameters are always identical to the ones used in the previous section, i.e., the time step and convergence tolerance are $2\times10^{-5}$\,s and $10^{-4}$, respectively.

The sensitivity to the time step length is negligible in the first phase of the response (up to strain $3.5\times10^{-3}$). The second phase, after the dynamic crack development, becomes very sensitive, especially the large time step gives quite different response because the simulation runs almost permanently without achieving the required converge limit (steps are accepted after 100 iterations regardless of the error). The second study shows that the reference convergence tolerance $10^{-4}$ is sufficient as it provides practically identical results as the tolerance $5\times10^{-5}$. However, twice or five times weaker tolerances exhibit some deviations, first in the second phase, but then also in the initial phase before dynamic crack development. The convergence criterion $5\times10^{-4}$ is clearly insufficient to deliver reliable results. 

Finally, the perturbation study demonstrates that even such a~complex simulation with a~large number of potential crack paths and failure modes exhibits surprising stability. Both tested $\eta$ values only lead to local modification of the response; the simulation quickly returns to its stable mode. The differences appear only in the second phase, after the dynamic event, which suffers from poor convergence.

\section{Conclusions}

This article presents a~systematic comparison of seven independent implementations of the Lattice Discrete Particle Model (LDPM), encompassing explicit and implicit time-integration strategies, transient and static solvers, and CPU- and GPU-based architectures. The implementations were developed independently by multiple research groups and were exercised on a~common set of benchmark problems of increasing mechanical and numerical complexity, ranging from linear elastic vibrations to highly nonlinear, fracture-dominated responses in tension and compression. The results provide a~unique, community-wide perspective on both the robustness of the LDPM formulation and the practical challenges associated with its numerical solution. 
        
With the exception of JAX-LDPM that runs on a~GPU architecture, all other implementations run on a~CPU architecture and allow the use of single and multiple threads. Three implementations (OAS, Project Chrono, and CAST3M) are open source; one implementation (FEMultiPhys) is provided as freeware software; and one implementation was pursued with user subroutines for a~widely used commercial software (ABAQUS). 

The first important contribution of this study is the demonstration of the critical role played by open-source or freeware implementations and openly distributed benchmark data. The availability of multiple open or freely accessible LDPM solvers, together with complete input files and reference results, enabled transparent cross-verification across platforms and numerical strategies. This openness was essential not only for identifying solver-specific strengths and limitations, but also for isolating issues related to time integration, convergence criteria, and implementation details rather than to the underlying constitutive model itself. Beyond the present comparison, the benchmark suite and datasets provided with this paper establish a~common reference framework that can support future LDPM developments, facilitate reproducibility, and lower the barrier for new researchers entering the field.

A second key outcome of this work is the recognition that defining objective and universally meaningful criteria for comparing different LDPM implementations is intrinsically difficult; more difficult than usual nonlinear finite element models due to local non-uniform response associated with LDPM's ability to capture material heterogeneity. While global quantities such as macroscopic stress–strain curves, energy balance, and computational time provide useful high-level indicators, they are often insufficient to fully characterize solver behavior in the presence of material heterogeneity, strain softening, and localization. Even when global responses are nearly identical, local fields, such as crack openings and fracture patterns, may exhibit nontrivial differences that depend on solver choices, convergence tolerances, mass matrix formulations, and numerical damping. Conversely, quantitative crack comparison metrics (e.g., correlation coefficients and NRMSE) may amplify small local discrepancies in simulations featuring pervasive cracking, making physically similar fracture patterns appear numerically dissimilar. These observations highlight that solver comparison for discrete mesoscale models cannot rely on a~single metric and must instead adopt a~multifaceted assessment combining global response, local fields, stability, and sensitivity analyses.

Furthermore, the results underscore the fundamental numerical complexity associated with simulating nonlinear behavior characterized by multiple interacting cracks and fragmentation, particularly in unconfined compression. In these regimes, the mechanical response transitions from quasi-static to strongly dynamic, with sudden energy releases, loss of solver convergence, and high sensitivity to numerical and solver-level parameters. Under these conditions, implicit solvers face limitations and they struggle to converge despite sophisticated algorithms. On the other hand explicit solvers do not suffer from loss of convergence and they outperform the implicit solvers in most of the presented examples even when computational time is adjusted for the used number of threads. 

However, the benefit of explicit solvers may be lost when simulating situations with realistic loading rates and for realistic load durations. For example, typical quasi-static concrete tests routinely conducted in concrete laboratories use a~loading rate 4 order of magnitude lower than the ones used in the unconfined compression examples presented in this article ($10^{-6}$\,sec$^{-1}$ compared to $10^{-2}$\,sec$^{-1}$) and, consequently, the actual test duration is in the order of minutes as opposed to fractions of a~second. Such a~different loading rate does not entail significant difference in the dynamics of the system since, as shown in this study, even for $10^{-2}$ sec$^{-1}$ the response is in the quasi-static regime. However in these cases the actual loading rate must be used in the simulations if the constitutive equations are rate-dependent (e.g. for the case of creep) and the conditional stability of explicit solvers becomes an~insurmountable challenge. 

An additional, more general implication of this study concerns the interpretation of numerical differences in the absence of an~exact reference solution. For highly nonlinear, heterogeneous material response such as that simulated by LDPM, an~analytical or ``exact'' solution is not available, and even experimental observations cannot uniquely provide global response and local fracture evolution. In this context, solver verification must rely on relative consistency rather than exact agreement. A~key result of this work is the demonstration that, across a~wide range of benchmarks and numerical strategies, all implementations produce globally and locally consistent responses that differ from one another by only a~few percent, even for the most challenging simulations, in terms of macroscopic quantities, energy measures, and crack pattern metrics. Such levels of discrepancy are small when viewed against the intrinsic randomness of LDPM response, arising from stochastic particle placement, geometric tessellation, and heterogeneous stress redistribution as reported in many past studies. When numerical differences among solvers fall below or within this inherent mesoscale variability, they should be regarded as numerically equivalent representations of the same physical response, rather than as meaningful deviations. This perspective is essential for the fair assessment of alternative numerical strategies and reinforces the conclusion that the observed differences primarily reflect unavoidable variability in discrete heterogeneous modeling, rather than deficiencies of any specific solver or implementation.

\section*{Data availability}
The data presented in the paper are available at  \href{https://doi.org/10.5281/zenodo.18197080}{10.5281/zenodo.18197080}. Detailed geometrical descriptions and geometry input files of all the benchmarks are available under the same link.

\section*{Acknowledgments}

The work of Erol Lale, Ke Yu, Bahar Ayhan, Giovanni Di Luzio, Matthew Troemner, and Gianluca Cusatis was partially supported by the Engineering Research and Development Center (ERDC) – Construction Engineering Research Laboratory (CERL) under Contract No. W9132T22C0015.

Jan Eliáš and Monika Středulová acknowledge financial support received from Czech Science Foundation under project No.~GA24-11845S. Monika also acknowledges Brno Ph.D. talent Scholarship funded by the Brno City Municipality used to implement LDPM material model to OAS. 

Julien Khoury and Gilles Pijaudier-Cabot acknowledge financial support from the investissement d’avenir French program (ANR-16-IDEX-0002) within the E2S Hub Newpores, the European Union’s Horizon 2020 research and innovation program EDENE under the Marie Skłodowska-Curie grant agreement No. 945416, and from the Communauté d’Agglomération Pau – Béarn – Pyrénées.

Tianju Xue and Jiawei Zhong acknowledge the support from the Young Collaborative Research Grant (YCRG) by the Research Grants Council of Hong Kong (Project No. C6002-24Y).

This research was supported in part through the computational resources and staff contributions provided for the Quest high performance computing facility at Northwestern University which is jointly supported by the Office of the Provost, the Office for Research, and Northwestern University Information Technology. 

\appendix

\section{Material parameters and constitutive model}
\label{appendix:material model}

\subsection{Fracturing behavior}
Fracturing behavior occurs when the normal strains are positive ($e_N > 0$). Tractions are obtained from effective traction, $t$, and effective strain, $e$, as follows
\parencite{Cusatis1}
\begin{align}
t_N&=\frac{t}{e}e_N & t_M&=\alpha \frac{t}{e} e_M & t_L=\alpha \frac{t}{e} e_L
\end{align} 

The effective traction $t=\sqrt{t_N^2+(t_M^2+t_L^2)/\alpha}$ is assumed to be incrementally elastic, $\dot{t}=E_0\dot{e}$, with the effective strain $e=\sqrt{e_N^2+\alpha(e_M^2+e_L^2)}$, and it should satisfy the inequality $0\leq t \leq\sigma_{bt}(e,\omega)$, where $\sigma_{bt}(e,\omega)$ is a~strain-dependent boundary imposed through a~vertical return algorithm and can be expressed as
\begin{align}
\sigma_{bt}(e,\omega) &= \sigma_{0}(\omega)\exp\left[-H_0(\omega)\frac{ \langle e_{\max}-e_0(\omega) \rangle }{\sigma_0(\omega)}\right]
\end{align} 
where the Macaulay brackets $\langle x \rangle= \max(x, 0)$ returns the positive part of the argument. This boundary $\sigma_{bt}$ undergoes exponential evolution in relation to the maximum effective strain,  $e_{\max}$, reached throughout the simulation. Coupling variable $\tan(\omega)=e_N/\sqrt{\alpha\left(e_M^2+e_L^2\right)}$ measures the degree of interaction between shear and normal loading. The variable $\sigma_0(\omega)$ is the strength limit for the effective traction and is defined as
\begin{align}
\sigma_0(\omega) = \sigma_t\frac{-\sin(\omega)+\sqrt{\sin(\omega)^2+4\alpha\cos(\omega)^2/r^2_{st}}}{2\alpha\cos(\omega)^2/r^2_{st}}
\end{align}
where $r_{st}$ is the ratio between the shear strength and the tensile strength, $r_{st} = \sigma_s/\sigma_t$
The exponential decay of the boundary $\sigma_{bt}$ initiates when the maximum effective strain reaches its elastic limit $e_0(\omega)=\sigma_0(\omega)/E_0$. The rate of decay is determined by the post-peak slope, also known as the softening modulus, which is assumed to follow a~power function of the internal variable $\omega$.
\begin{align}
H_0(\omega) = \frac{H_s}{\alpha}+\left(H_t- \frac{H_s}{\alpha}\right)\left(\frac{2\omega}{\pi}\right)^{n_t}
\end{align} 
where $H_t = 2E_0/(l_t/l-1)$ and $l_t = 2E_0G_t/\sigma^2_t$. $G_t$ is the fracture energy, and $l$ is the length of the tetrahedron edge corresponding to the current facet. $H_s=r_sE_0$. In most cases, $r_s=0$. Figure~\ref{fig:LDPM2}e reports examples of traction vs strain curves according to these constitutive equations. 

During unloading the effective traction undergoes elastic reduction with effective strain until it reaches zero, then it maintains zero value with a~further decrease in effective strain. During reloading, the effective traction remains zero until the effective strain reaches the reloading strain limit $e_{\mathrm{tr}} = k_t(e_{\max}-\sigma_{bt}/E_0)$. Beyond this threshold, the effective traction elastically increases. Material parameters $k_t$ governs the size of hysteresis cycles and the dissipated energy. 

\subsection{Compressive behavior}
For compressive behavior ($e_N < 0$), the normal traction satisfies the inequality $-\sigma_{bc}(e_D,e_V)\leq t_N\leq0$. The normal traction $t_N$ is assumed to  be incrementally elastic within the boundary, $\dot{t}_N=E_{Nc}\dot{e}_N$. $E_{Nc}$ is the loading-unloading stiffness that increases during unloading. 
\begin{align}
E_{Nc} = \begin{cases}
E_0 & -t_N\leq\sigma_{c0} \\
E_d & \text{otherwise}
\end{cases}
\end{align}

The compressive boundary can be expressed as
\begin{align}
\sigma_{bc}(e_D, e_V) &= 
\begin{cases}
\sigma_{c0} & -e_{DV}\leq 0 
\\
\sigma_{c0} + \left\langle -e_{DV}-e_{c0} \right\rangle H_c\left(r_{DV}\right) & 0\leq-e_{DV}\leq e_{c1} 
\\
\sigma_{c1}(r_{DV})\exp\left[(-e_{DV}-e_{c1})\dfrac{H_c(r_{DV})}{\sigma_{c1}(r_{DV})}\right] & \text{otherwise} 
\end{cases}
\end{align}
where $e_{DV}=e_V+\beta e_D$ ($\beta$ is a~material parameter, $e_V$ is the volumetric strain from the tetrahedron volume and $e_D = e_N - e_V$ is the deviatoric stress) and $e_{c0}=\sigma_{c0}/E_0$ is
the compaction strain at the beginning of the pore collapse, $e_{c1}=\kappa_{c0}e_{c0}$ is the compaction strain at which rehardening begins, $\kappa_{c0}$ is the material parameter governing the rehardening and $\sigma_{c1}(r_{DV}) = \sigma_{c0}+(e_{c1}-e_{c0})H_c(r_{DV})$.
$H_c(r_{DV})$ is the initial hardening modulus
\begin{align}
H_c(r_{DV}) = \frac{H_{c0}-H_{c1}}{1+\kappa_{c2}\langle r_{DV}-\kappa_{c1}\rangle}+H_{c1}
\end{align}
and $r_{DV}$ is deviatoric to volumetric strain ratio
\begin{align}
r_{DV} = \begin{cases}
-\dfrac{|e_D|}{e_V-e_{V0}}, & e_{V}\leq0 \\[2mm]
\dfrac{|e_D|}{e_{V0}}, & e_{V} > 0\\
\end{cases}
\end{align}
where $e_{V0} = \kappa_{c3} \sigma_{c0} / E_0$.

Figure~\ref{fig:LDPM2}f reports examples of traction vs strain curves according to these constitutive equations.

\subsection{Frictional behavior}
In the compression case, the shear strength increases due to frictional effect. The frictional phenomena is simulated by classical incremental plasticity, while the incremental traction can be defined as $\dot{\sigma}_M=E_0\alpha(\dot{e}_M-\dot{e}^p_M)$, $\dot{\sigma}_L=E_0\alpha(\dot{e}_L-\dot{e}^p_L)$. The plastic strain increments are assumed to obey the normality rule $\dot{e}^p_M=\dot{\lambda}\partial\phi/\partial t_M$, $\dot{e}^p_L=\dot{\lambda}\partial\phi/\partial t_L$. The plastic potential can be expressed as $\phi=\sqrt{t^2_M+t^2_L}-\sigma_{bs}(t_N)$, where the shear strength $\sigma_{bs}$ is calculated as
\begin{align}
\sigma_{bs}(\sigma_N) = \sigma_s+(\mu_0-\mu_{\infty})\sigma_{N0}-\mu_{\infty} t_N-(\mu_0-\mu_{\infty})\sigma_{N0}\exp(t_N/\sigma_{N0})
\end{align} 
where $\sigma_s$ is the cohesion (shear strength), $\mu_0$ and $\mu_{\infty}$ are the initial and final friction coefficients, and $\sigma_{N0}$ is the normal traction at which the friction coefficient transitions from $\mu_0$ to $\mu_{\infty}$. The cohesion $\sigma_s$ would degrade when shear softening modulus $H_s\neq0$ (softening in pure shear), where $H_s=r_sE_0$. In most cases, $r_s=0$.

\begin{table} [tb!]
    \centering
    \begin{tabular}{ llll } 
        \toprule
        \textbf{parameter} &\textbf{symbol} &\textbf{value} & \textbf{unit} \\
        \midrule
        Density & $\rho$ & 2\,380 & kg/m$^3$ \\ 
        Normal Modulus & $E_0$ & 60\,273 & MPa \\ 
        Alpha &$\alpha$ & 0.25 & \\ 
        Tensile Strength & $\sigma_t$ & 3.44 & MPa \\ 
        Tensile Characteristic Length & $l_t$ & 500 & mm \\ 
        Fracture energy & $G_t$ or ($l_t\sigma_t^2/(2E_0)$) & 0.0491 & N/mm \\ 
        Shear Strength Ratio & $r_t$ or ($\sigma_s$/$\sigma_t$) & 2.6 &  \\ 
        Softening Exponent & $n_t$ & 0.4 &  \\ 
        Compressive Yielding Strength & $\sigma_{c0}$ & 150 & MPa \\ 
        Initial Hardening Modulus Ratio & $H_{c0}/E_0$ & 0.40 &  \\ 
        Transitional Strain Ratio & $\kappa_{c0}$ & 4 &  \\ 
        Deviatoric Strain Threshold Ratio & $\kappa_{c1}$ & 1 &  \\ 
        Deviatoric Damage Parameter & $\kappa_{c2}$ & 5 &  \\ 
       Volumetric Strain Parameter & $\kappa_{c3}$ & 0.1 &  \\ 
        Initial Friction & $\mu_{0}$ & 0.4 &  \\ 
        Asymptotic Friction & $\mu_{\infty}$ & 0 &  \\ 
        Transitional Stress & $\sigma_{N0}$ & 600 & MPa \\ 
        Densification Ratio & $E_d/E_0$ & 1 & \\ 
        Volumetric Deviatoric Coupling & $\beta$ & 0 & \\ 
        Tensile Unloading & $k_t$ & 0 & \\ 
        Shear Unloading & $k_s$ & 0 & \\ 
        Compressive Unloading & $k_c$ & 0 & \\ 
        Shear Softening Modulus Ratio & $r_s$ & 0 & \\ 
        Final Hardening Modulus Ratio & $H_{c1}/E_0$ & 0.1 &  \\ 
        \bottomrule
    \end{tabular}
    \caption{Material parameters of LDPM used throughout the paper.}
    \label{t:matparams}
\end{table}

\printbibliography
\end{document}